\documentclass[a4paper,11pt]{article}
\pdfoutput=1 
\usepackage{jheppub} 
\usepackage[T1]{fontenc} 
\usepackage{graphicx}
\graphicspath{{fig/}} \DeclareGraphicsExtensions{.pdf}
\usepackage{amsmath}
\usepackage{bm}
\usepackage{slashed}
\usepackage{amsfonts}

\newcommand{\jpsi}{{J/\psi}}
\newcommand\psip{{\psi^\prime}}

\newcommand{\chicj}{\chi_{cJ}}

\newcommand{\states}[3]{{^#1\hspace{-0.6mm}#2_{#3}}}
\newcommand{\state}[4]{{^#1\hspace{-0.6mm}#2_{#3}^{[#4]}}}

\newcommand\cSa{\states{3}{S}{1}}
\newcommand\aPa{\states{1}{P}{1}}
\newcommand\cPj{\states{3}{P}{J}}

\newcommand\CSaSz{\state{1}{S}{0}{1}}
\newcommand\CSaPa{\state{1}{P}{1}{1}}
\newcommand\CScSa{\state{3}{S}{1}{1}}
\newcommand\CScPz{\state{3}{P}{0}{1}}
\newcommand\CScPa{\state{3}{P}{1}{1}}
\newcommand\CScPb{\state{3}{P}{2}{1}}

\newcommand\COaSz{\state{1}{S}{0}{8}}
\newcommand\COaPa{\state{1}{P}{1}{8}}
\newcommand\COcSa{\state{3}{S}{1}{8}}
\newcommand\COcPz{\state{3}{P}{0}{8}}
\newcommand\COcPa{\state{3}{P}{1}{8}}
\newcommand\COcPb{\state{3}{P}{2}{8}}
\newcommand\COcPj{\state{3}{P}{J}{8}}

\newcommand\mo{{\mathcal O}}

\newcommand{\LDME}[2]{\langle\mo^{#1}(#2)\rangle}
\newcommand\mops{\LDME{\jpsi}{\CScSa}}
\newcommand\mopa{\LDME{\jpsi}{\COaSz}}
\newcommand\mopb{\LDME{\jpsi}{\COcSa}}

\newcommand\mopj{\LDME{\jpsi}{\COcPj}}
\newcommand\mopjz{\LDME{\jpsi}{\COcPz}}

\newcommand{\vt}[1]{{{\boldsymbol #1}_\perp}}
\newcommand{\vtn}[2]{{{\boldsymbol #1}_{#2\perp}}}
\newcommand{\vb}{{\vt{b}}}
\newcommand{\vx}{{\vt{x}}}
\newcommand{\vy}{{\vt{y}}}
\newcommand{\vz}{{\vt{z}}}
\newcommand{\vr}{{\vt{r}}}
\newcommand{\vD}{{\vt{\Delta}}}
\newcommand{\vl}{{\vt{l}}}
\newcommand{\vp}{{\vt{p}}}
\newcommand{\vq}{{\vt{q}}}
\newcommand{\vk}{{\vt{k}}}
\newcommand{\vka}{{\vtn{k}{1}}}

\newcommand{\vrz}{{\vtn{r}{0}}}

\newcommand{\vtp}[1]{{{\boldsymbol #1}'_\perp}}
\newcommand{\vtnp}[2]{{{\boldsymbol #1}'_{#2\perp}}}

\newcommand{\vxp}{{\vtp{x}}}
\newcommand{\vyp}{{\vtp{y}}}
\newcommand{\vrp}{{\vtp{r}}}
\newcommand{\vlp}{{\vtp{l}}}

\newcommand{\vkp}{{\vtp{k}}}
\newcommand{\vkap}{{\vtnp{k}{1}}}

\newcommand{\x}{\vx}
\newcommand{\q}{\vq}
\newcommand{\p}{\vp}
\renewcommand{\b}{\vb}
\newcommand{\xt}{\vx}
\newcommand{\yt}{\vy}
\newcommand{\zt}{\vz}

\newcommand{\nc}{N_c}

\title{\boldmath Quarkonium production in high energy proton-nucleus collisions: CGC meets NRQCD}

\author[a]{Zhong-Bo Kang,}
\author[b]{Yan-Qing Ma,}
\author[b]{and Raju Venugopalan}

\affiliation[a]{Theoretical Division,
                   Los Alamos National Laboratory,
                   Los Alamos, NM 87545, USA}
\affiliation[b]{Physics Department,
                   Brookhaven National Laboratory,
                   Upton, NY 11973, USA}

\emailAdd{zkang@lanl.gov}
\emailAdd{yqma@bnl.gov}
\emailAdd{raju@bnl.gov}

\date{\today}

\abstract{
We study the production of heavy quarkonium states in high energy
proton-nucleus collisions. Following earlier work of Blaizot, Fujii,
Gelis, and Venugopalan, we systematically include both small $x$
evolution and multiple scattering effects on heavy quark pair
production within the Color Glass Condensate (CGC) framework. We
obtain for the first time expressions in the Non-Relativistic QCD
(NRQCD) factorization formalism for heavy quarkonium differential
cross sections as a function of transverse momentum and rapidity. We
observe that the production of color singlet heavy quark pairs is
sensitive to both ``quadrupole" and ``dipole" Wilson line
correlators, whose energy evolution is described by the
Balitsky-JIMWLK equations. In contrast, the color octet channel is
sensitive to dipole correlators alone. In a quasi-classical
approximation, our results for the color singlet channel reduce to
those of Dominguez et. al.~\cite{Dominguez:2011cy}. We compare our
results to those obtained combining the CGC with the color
evaporation model and point to qualitative differences in the two
approaches.
}

\begin{document}
\maketitle
\flushbottom

\section{Introduction}\label{sec:intro}

Quarkonium production in proton-nucleus collisions provides an
excellent laboratory for studying the interaction of colored heavy
quark probes with an extended colored medium. The large mass scale
provided by the heavy quarks suggests that their interactions can be computed systematically in a weak coupling
framework. However, the use of heavy quarks as a probe of colored
media has been bedeviled by the complexities encountered in
understanding the production of heavy quark states in more
elementary collisions. The development of the Non-Relativistic QCD
(NRQCD) framework~\cite{Bodwin:1994jh} provided a systematic power
counting to organize this complexity, and there has been a
tremendous amount of work since in making this a quantitative
framework--for recent summaries of the state of the art, see for
example~\cite{Brambilla:2010cs,Chao:2012upa,Bodwin:2013nua}.
Specifically, we should point to recent next-to-leading order
studies which find that the yield of all quarkonia states in
proton-proton collisions can be described in NRQCD factorization,
including the $\jpsi$~\cite{Ma:2010yw,Butenschoen:2010rq},
$\psip$\cite{Ma:2010jj}, $\chicj$~\cite{Ma:2010vd} and $\Upsilon
(nS)$~\cite{Wang:2012is,Gong:2013qka} states.

At the same time, a systematic weak coupling framework, the Color
Glass Condensate (CGC), was developed to describe the high parton
density effects of small $x$ QCD evolution and coherent multiple
scattering~\cite{Kovchegov:2012mbw,Gelis:2010nm,Weigert:2005us,Iancu:2003xm}.
At high energies, the typical momentum transfer from partons in the
medium to the probe is no longer soft and is characterized by a
semi-hard ``saturation" scale  $Q_{s}^2 \gg \Lambda_{\rm QCD}^2$.
This
scale~\cite{Gribov:1984tu,Mueller:1985wy,McLerran:1993ni,McLerran:1993ka}
separates highly occupied gluon transverse momentum modes from
perturbative dynamics at large transverse momentum. The saturation
scale is dynamically generated from the fundamental scale of the
theory; it is proportional to the density of partons in the
transverse radius of the nucleus, and grows with energy. Because the
running of the coupling is controlled by this scale, asymptotic
freedom tells us that the coupling of the colored partonic probe
should be weak and will become weaker at higher energies. The hope
therefore is that with some effort one can compute systematically
the many-body structure of hadrons and nuclei at high energies.

In particular, the  CGC has been widely applied to study a number of
final states in proton-nucleus collisions-for reviews,
see~\cite{Albacete:2013tpa,JalilianMarian:2005jf}. For other
approaches to quarkonium production in proton-nucleus collisions,
see
\cite{Albacete:2013ei,Vogt:2010aa,McGlinchey:2012bp,Arleo:2013zua,Sharma:2012dy,Kopeliovich:2013yfa}.
An attractive feature of the CGC effective theory is that one can
quantify what one means by dilute or dense scatterers as a function
of energy and mass number~\cite{Blaizot:2004wu}. Typically in
proton-nucleus collisions we encounter a ``dilute-dense" system. To
be more precise, the ``dilute" limit is a systematic expansion of
amplitudes to lowest order in the ratio of the saturation momentum
of the proton to the typical transverse momentum exchanged by the
proton in the reaction ($Q_{s,p}/k_{\perp,p} \ll 1$). In turn,
the ``dense" limit corresponds to keeping in the amplitude all orders
in the ratio of the saturation momentum of the nucleus relative to
the momentum exchanged by the nucleus ($Q_{s,A}/k_{\perp,A} \sim
1$).
At very high energies, the power counting in proton-nucleus collisions may be closer
to that in proton-proton collisions. Further, at rapidities far from the proton beam, the power counting in proton-nucleus collisions may be closer to that in nucleus-nucleus collisions.

Quarkonium pair production was first studied in the CGC framework in
the limit of small $x$ and large transverse
momentum~\cite{Gelis:2003vh}. It was shown explicitly that in this
limit one recovers the $k_\perp$-factorization results\footnote{All
these results differ in detail from a similar result obtained at the
same time in~\cite{Levin:1991ry}.}  of Collins and
Ellis~\cite{Collins:1991ty} and Catani, Ciafaloni and
Hautmann~\cite{Catani:1990eg}.  However, for $k_\perp \leq Q_{s}$,
it was shown\footnote{Here, and henceforth, we will use $Q_s$ to
denote the saturation scale in the nucleus.} in
\cite{Blaizot:2004wv} that $k_\perp$-factorization is broken
explicitly in quark pair production, even at leading order in
proton-nucleus collisions\footnote{A closely related computation was
carried out in \cite{Tuchin:2004rb}. The results of
\cite{Blaizot:2004wv} were independently confirmed in a different
approach, which focused on the effect on single spin asymmetries on
heavy quark pair production~\cite{Akcakaya:2012si}. A computation
that extends the work of \cite{Blaizot:2004wv} to include rapidity
evolution between the heavy quarks can be found in
\cite{Kovchegov:2006qn}.}. The magnitude of the breaking of
$k_\perp$-factorization for single inclusive quark production and
quark pair production was quantified respectively in
\cite{Fujii:2005vj} and \cite{Fujii:2006ab}.

The results in these papers were derived for heavy quark pair
production but the projection of these results for specific
quarkonium states were not considered. In the same general
framework, $J/\psi$ production from quark pairs in color singlet and
color octet configurations were previously considered in
\cite{Kharzeev:2005zr,Kharzeev:2008nw,Kharzeev:2008cv,Dominguez:2011cy,Kharzeev:2012py}.
However, these derivations were performed in a quasi-classical
approximation, and the effects of QCD evolution were only included
heuristically through energy evolution of the saturation scale. The
formalism for heavy quark pair production developed in
\cite{Blaizot:2004wv,Fujii:2006ab} was recently combined with the
color evaporation model to compute $J/\psi$ and $\Upsilon$
production in high energy proton-nucleus
collisions~\cite{Fujii:2013gxa}.

In this paper, we project the amplitude for heavy quark pairs
computed in \cite{Blaizot:2004wv} on to color singlet and color
octet configurations. Interestingly, the energy/rapidity evolution
of the corresponding short-distance cross-sections, as we shall
discuss further, is described by different combinations of
multi-gluon correlators in the CGC framework. These short distance
cross-sections are matched on to long distance vacuum NRQCD matrix
elements to provide detailed expressions for the cross-sections for
all common S and P wave quarkonium states in proton-nucleus
collisions\footnote{In very high energy proton-nucleus collisions,
at small $x$, the hadronization of heavy quark pairs into quarkonium
states happens well after the collision. It is therefore reasonable
to expect that the vacuum NRQCD matrix elements accurately represent
the hadronization physics in these collisions.}. In a follow up
paper, we will compare our results to data on quarkonium production
in deuteron-nucleus collisions at RHIC and proton-nucleus collisions
at the LHC. The large amount of data now available at different
energies, and for a variety of quarkonium states promises to provide
sensitive tests of both the CGC and and the NRQCD formalisms.

The paper is organized as follows. In section~\ref{sec:amp}, we
provide a brief recap of the CGC framework and key results for heavy
quark pair production. In section~\ref{sec:crosssection}, we discuss
the matching of these results to the NRQCD formalism. We describe
simplifications of our results that occur in the limit of large
$N_c$, the collinear limit, and at high $p_\perp$ of the quarkonium.
A comparison of our results to previous results obtained in the
quasi-classical approximation is presented in
section~\ref{sec:comparison}. In this section, we also compare our
results to results obtained by combining the CGC framework with the
Color Evaporation model (CEM). We end with a brief summary and
outlook on ongoing work. Some essential details of the computations
are presented in two appendices.

\section{Quark pair production in the Color Glass Condensate}
\label{sec:amp}

\subsection{General discussion}
In the CGC formalism, the proton-nucleus collision is described as a
collision of two classical fields originating from color sources
representing the large $x$ degrees of freedom in the proton and the
nucleus.  The color source distribution generating
the classical field in each projectile is evolved from
initial valence distribution at large $x$ to the rapidity of interest in the collision. The
gauge fields of gluons produced in the collision are determined by
solving the Yang-Mills equations
\begin{equation}
[D_\mu,F^{\mu\nu}] = J^\nu \, .
\label{eq:YM}
\end{equation}
Here $J^\nu$ is the color current of the sources, which can be expressed
at leading order in the sources as
\begin{equation}
J_a^\nu = g\delta^{\nu+}\delta(x^-)\,\rho_{p,a}(\x_\perp) + g\delta^{\nu-}\delta(x^+)\rho_{_A,a}(\x_\perp) \, ,
\label{eq:current}
\end{equation}
where $\rho_p$ is the number density of ``valence" partons in the
proton moving in the $+z$ direction at the speed of light.  Likewise,
$\rho_{_A}$ is the number density of ``valence" partons in the nucleus
moving in the opposite light cone direction. To solve these equations, one needs to impose a gauge fixing condition.
Further, covariant current conservation requires that
\begin{equation}
[D_\nu,J^\nu]=0\; .
\label{eq:cur-conv}
\end{equation}
The latter equation in general implies that eq.~(\ref{eq:current})
for the current receives corrections that are of higher order in the
sources $\rho_p$ and $\rho_{_A}$, because of the radiated field. The
solution of eqs.~(\ref{eq:YM}), (\ref{eq:current}) and
(\ref{eq:cur-conv}) has been determined to all orders in both
sources only
numerically~\cite{Krasnitz:1998ns,Krasnitz:1999wc,Krasnitz:2003jw,Lappi:2003bi}.
To lowest order in the proton source (as appropriate for a dilute
proton source) and to all orders in the nuclear source, analytical
results are available and an explicit expression for the gauge field
to this order, in Lorentz gauge, is given\footnote{The expression
for the gauge field was also obtained in \cite{Gelis:2005pt} in the
light-cone gauge of the proton, and in \cite{Dumitru:2001ux} in
Fock-Schwinger gauge $x^+A^-+x^-A^+=0$.} in
ref.~\cite{Blaizot:2004wu}. The amplitude for pair production to
this order is obtained by computing the quark propagator in the
background corresponding to this gauge field~\cite{Blaizot:2004wv}.

The probability for producing a single $q\bar{q}$ pair for a given
distribution of color sources ($\rho_p$ in the proton and $\rho_A$
in the nucleus)  is
\begin{equation}
P_1[\rho_p,\rho_{_A}]=
\int\frac{d^3\q}{(2\pi)^3 2E_\q}
\int\frac{d^3\p}{(2\pi)^3 2E_\p}
\left|{\cal M}_{_{F}}(\q,\p)\right|^2\; ,
\label{eq:P1-def}
\end{equation}
where ${\cal M}_{_{F}}(\q,\p)$ is the amputated time-ordered quark
propagator in the presence of the classical field generated by the sources.  The expression, as it stands, is not
gauge invariant. To convert this probability into a physical cross-section, we first average over the
initial classical sources $\rho_p$ and $\rho_{_{A}}$ respectively with
the weights $W_p[x_p,\rho_p]$ and $W_{_{A}}[x_{_A},\rho_{_{A}}]$. These weight functionals are gauge invariant by construction.
 We subsequently integrate over all impact parameters $\b$, to obtain the cross section to produce a heavy quark pair:
\begin{equation}
\sigma=\int d^2\b \int[D\rho_p][D\rho_{_{A}}]
\,W_p[x_p,\rho_p]\,W_{_{A}}[x_{_A},\rho_{_{A}}]
P_1[\rho_p,\rho_{_A}]\; .
\label{eq:source-avg}
\end{equation}
This formula incorporates both multiple scattering effects and those
of the small $x$ quantum evolution. The multiple scattering effects
are included in i) the classical field obtained from solving the
Yang-Mills equation in eq.~(\ref{eq:YM}) with the current in
eq.~(\ref{eq:current}), ii) in the propagator of the quark in this
classical field, as well as iii) in the small $x$ renormalization
group evolution of the color source distribution of the nucleus.

The leading logarithmic small $x$ evolution is included in the
evolution of the weight functionals, $W_p$ and $W_{_{A}}$, of the
target and projectile with $x$.  The arguments $x_p$ and $x_{_A}$
denote the scale in $x$ separating the large-$x$ static sources from
the small-$x$ dynamical fields. In the McLerran-Venugopalan
model~\cite{McLerran:1993ni,McLerran:1993ka}, the functional
$W_{_{A}}$ that describes the distribution of color sources in the
nucleus is a Gaussian in the color charge density\footnote{This is
true modulo terms parametrically suppressed in
$A$~\cite{Jeon:2004rk,Jeon:2005cf,Dumitru:2011ax}. Note further that
in the CGC framework, the saturation scales enter through this
initial condition.} in $\rho_{_{A}}$. A Gaussian distribution of
sources is equivalent to the QCD Glauber model of independent
multiple scattering~\cite{Blaizot:2004wu}. We shall address this
point further later in our discussion of the quasi-classical limit
of quarkonium production.  In general, however, this Gaussian
distribution of color sources is best interpreted as the initial
condition for a non-trivial evolution of
$W_{_{A}}[x_{_A},\rho_{_{A}}]$ with $x_{_A}$. The evolution of the
$W$'s is described by a Wilsonian renormalization group equation,
the JIMWLK equation; the corresponding hierarchy of equations for
expectation values of multi-gluon is called the Balitsky-JIMWLK
hierarcy~\cite{Balitsky:1995ub,JalilianMarian:1997dw,Iancu:2000hn}.
We will discuss the Balitsky-JIMWLK hierarchy further in the
following section.

\subsection{Heavy quark pair
production amplitude}\label{sec:amplitude}
For our purpose here, the relevant quantity is the heavy quark pair production amplitude computed in \cite{Blaizot:2004wv}.
We begin with the kinematic notations for the process\footnote{Note that these differ slightly from \cite{Blaizot:2004wv}.}
\begin{align}
p(p_p) + A(p_A) \to Q\left(\frac{p}{2}+q\right)\bar{Q}\left(\frac{p}{2}-q\right)+X \, .
\end{align}
We will assume that the proton moves in the $+z$ direction with momentum $p_p=(p_p^+, 0^-, 0_\perp)$ and the nucleus
in the $-z$ direction with momentum $p_A=(0^+, p_A^-, 0_\perp)$.
Here $p$ and $q$ correspond respectively to  the total momentum of the heavy quark pair and one half of the relative momentum of the quark and anti-quark constituting the pair. The on-shell constraints on the quark and the anti-quark
$\left(p/2+q\right)^2=m^2$ and $\left(p/2-q\right)^2=m^2$ imply that
\begin{align}
p\cdot q =0\quad \text{and} \quad p^2=4(m^2-q^2)\;,
\end{align}
with $m$ the heavy quark mass.

\begin{figure}[htb!]
 \begin{center}
 \includegraphics[width=0.90\textwidth]{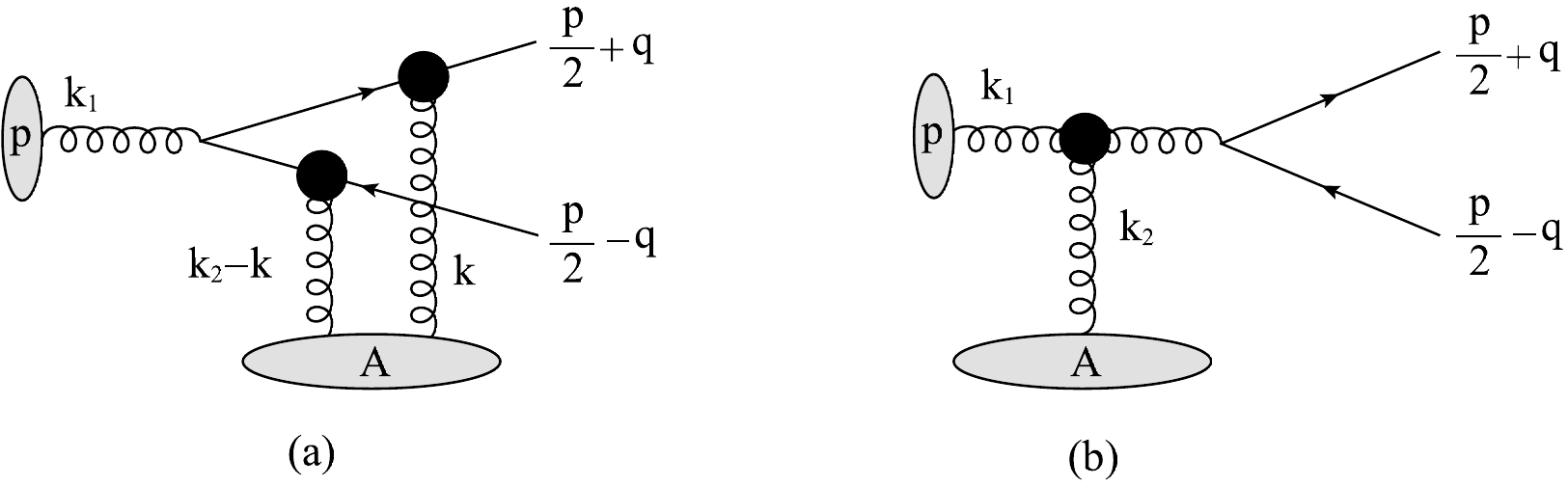}
 \caption[]{Feynman diagram representation of heavy quark pair production in pA
 collisions. The two diagrams represent respectively the two terms in
 eq.~\eqref{eq:QQamp}, where the black dots denote the Wilson lines that resum all the multiple
 scatterings of either the associated gluon or the heavy quark pair off
 the color field of the nucleus.
   \label{fig:feyndiag}}
 \end{center}
\end{figure}
Within the CGC formalism, the amplitude to produce a heavy quark
pair has two contributions. One of these, illustrated in
fig.\ref{fig:feyndiag}\,(a), is where a gluon from the proton emits
a heavy quark pair before the collision with the target, while the
other, illustrated in fig.\ref{fig:feyndiag}\,(b), is where the
gluon emits the heavy quark pair after the collision with the
target~\cite{Blaizot:2004wv}. We denote $k_1=(x_p p_p^+,0,\vka)$ as
the momentum of the gluon from the proton, $k_2=p-k_1=(0, x_A
p_A^-,\vp-\vka)$ as the total momentum of gluons from the nucleus,
and $\rho_{p}$ and $\rho_{A}$ as the densities of color sources in
the proton and nucleus, respectively. The heavy quark pair
production amplitude then reads~\cite{Blaizot:2004wv}
\begin{align}\label{eq:QQamp}
\begin{split}
&M^F_{s \bar{s}; i \bar{i}}(p,q)=\frac{g_s^2}{(2\pi)^4}
\underset{\vka,\vk}{\int} \frac{\rho_{p,a}(x_p,\vka)}{k_{1\perp}^2}
\underset{\vx, \vy}\int e^{i
\vk\cdot\vx} e^{i (\vp-\vk-\vka)\cdot\vy}\\
&\times \bar{u}_{s; i}\left(\frac{p}{2}+q\right) \left[
T_{q\bar{q}}\left(p,q,\vka,\vk\right)V_F(\vx)t^a V_F^\dagger(\vy) +
T_g(p,\vka)t^b V_A^{b a}(\vx)\right] v_{\bar{s};
\bar{i}}\left(\frac{p}{2}-q\right),
\end{split}
\end{align}
where $s$ and $i$ ($\bar{s}$ and $\bar{i}$) are spin index and color
index of quark (antiquark), respectively, and
$\int_\vk\equiv\int {d^2\vk}$, $\int_\vx\equiv \int {d^2 \vx}$. The
functions $T_{q\bar{q}}\left(p,q,\vka,\vk\right)$ and $T_g(p,\vka)$
are defined to be
\begin{subequations}
\begin{align}
\begin{split}
&T_{q\bar{q}}\left(p,q,\vka,\vk\right)\\
\equiv &\frac{\gamma^+ \left( \frac{\slashed{p}}{2} + \slashed{q}
-\slashed{k} +m\right)\gamma^- \left( \frac{\slashed{p}}{2} +
\slashed{q} -\slashed{k} -\slashed{k}_1+m\right)\gamma^+}
{2\left(\frac{p^+}{2}-q^+\right)\left[\left(\frac{\vp}{2}+\vq-\vk\right)^2+m^2\right]
+2\left(\frac{p^+}{2}+q^+\right)\left[\left(\frac{\vp}{2}+\vq-\vk-\vka\right)^2+m^2\right]},
\end{split}\\
\begin{split}
T_g(p,\vka)\equiv\frac{\slashed{C}_L(p,\vka)}{p^2},
\end{split}
\end{align}
\end{subequations}
with $C^\mu_L(p,\vka)$ the well-known Lipatov effective vertex,
\begin{subequations}
\begin{align}
C^+_L(p,\vka)=&-\frac{k_{1\perp}^2}{p^-}+p^+,\\
C^-_L(p,\vka)=&\frac{\left(\vp-\vka\right)^2}{p^+}-p^-,\\
C^i_L(p,\vka)=&-2k_1^i+p^i.
\end{align}
\end{subequations}
The Wilson lines  $V_F(\vx)$ and $V_A(\vx)$ are
defined as
\begin{subequations}
\begin{align}
\label{eq:target-Wilsonlines}
V_F(\vx)\equiv &\mathcal{P}_+ \text{exp}\left[-i
g_s^2\int_{-\infty}^\infty d
z^+\frac{1}{{\boldsymbol\nabla}_\perp^2}\rho_A(z^+,\vx)\cdot
t\right],\\
V_A(\vx)\equiv &\mathcal{P}_+ \text{exp}\left[-i
g_s^2\int_{-\infty}^\infty d
z^+\frac{1}{{\boldsymbol\nabla}_\perp^2}\rho_A(z^+,\vx)\cdot
T\right],
\end{align}
\end{subequations}
where $\mathcal{P}_+$ denotes the ``time
ordering" along the $z^+$ axis, and $t^a$ ($T^a$) are the $SU(N_c)$ generators of the fundamental
(adjoint) representation.

We note that the amplitude in eq.~(\ref{eq:QQamp}) agrees exactly
with the $k_\perp$-factorized result derived in \cite{Gelis:2003vh}
when the Wilson line correlators are expanded to first order in
$\rho_A/{\boldsymbol\nabla}_\perp^2$. In general, however,
$k_\perp$-factorization is explicitly broken for pair production in
proton-nucleus collisions\footnote{This is to be contrasted to the
result, shown by several authors, that $k_\perp$-factorization holds
at leading order for single inclusive gluon production in
proton-nucleus collisions.}.

\section{Quarkonium production cross section}\label{sec:crosssection}

In this section, we will discuss the matching of the results of the
previous section to the NRQCD formalism. We will derive explicit
expressions for the short distance cross-sections, and the
associated small $x$ multi-gluon correlators in the large $N_c$
limit.  We shall also discuss the limit when the transverse momentum
of the gluon exchanged by the proton is small, and demonstrate that
collinear factorization is recovered on the proton side to leading
order. Finally, we will discuss the power counting of the color
singlet and color octet channels in the large $p_\perp$ limit of our
computation.

\subsection{Quarkonium production within the NRQCD factorization
formalism}\label{sec:NRQCD}

We begin with a brief review of the NRQCD factorization
formalism~\cite{Bodwin:1994jh}. The inclusive production of a heavy
quarkonium state $H$ in the process $p+A\to H+X$ is expressed in
this framework as
\begin{align}\label{eq:factorization}
d\sigma_H=\sum_\kappa d \hat{\sigma}^\kappa
\langle\mo^{H}_\kappa\rangle.
\end{align}
Here $\kappa=\state{{2S+1}}{L}{J}{C}$ are the quantum numbers of the
produced intermediate heavy quark pair, where $S$, $L$ and $J$ are
the spin, orbital angular momentum and total angular momentum,
respectively. The symbol $C$ here denotes the color state of the
pair, which can be either color singlet (CS) with $C=1$ or color
octet (CO) with $C=8$. In eq.~(\ref{eq:factorization}), $d
\hat{\sigma}^\kappa$ are the short distance
coefficients\footnote{Readers should note that these coefficients
for different channels have differing mass dimensions, as do of
course then the long distance matrix elements.} for the production
of a heavy quark pair with quantum numbers $\kappa$. These can be
calculated perturbatively and can be factorized from the
non-perturbative NRQCD long distance matrix elements
(LDME)\footnote{The $S$-wave LDMEs have mass dimension of $[M]^3$
while $P$-wave LDMEs have mass dimension of $[M]^5$. Further, for
our convenience we shall use a definition for CS LDMEs
\cite{Petrelli:1997ge}, which is different from the original BBL
convention~\cite{Bodwin:1994jh} by a factor of $1/(2N_c)$. For
example,
\begin{align}\label{eq:O3s11}
\mops=\frac{1}{2N_c}\mops_{\text
{BBL}}=\frac{3}{4\pi}\left|R(0)\right|^2 \left[1+O(v^4)\right],
\end{align}
where $R(0)$ is the $\jpsi$ wavefunction at the origin. }
$\langle\mo^{H}_\kappa\rangle$. Specifically, the LDMEs  describe
the hadronization of a heavy quark pair with quantum numbers
$\kappa$ to the quarkonium state $H$. They are universal and can be
determined by fitting experimental data~\cite{Brambilla:2010cs}. The
LDMEs are organized by powers of $v$, the relative velocity of heavy
quark pair in the heavy quarkonium bound state. As $v$ is a small
non-relativistic velocity in the heavy quarkonium system, one needs
only a few LDMEs in practice.

For example, there are four independent LDMEs which are important
for phenomenological study of $\jpsi$ production\footnote{The
magnitude of the CS LDME $\mops$ is largest in powers in $v$, while
the three CO LDMEs listed in eq.~\eqref{eq:states} are relatively
power suppressed by $v^3$, $v^4$ and $v^4$, respectively. For
$\jpsi$ production with a large transverse momentum $p_\perp$ at
hadron colliders, one finds that the contribution of the CS channel
at leading order in $\alpha_s$ is suppressed by $m^2/p_\perp^2$
compared to the $\COaSz$ and $\COcPj$ channels, and even further
suppressed by $m^4/p_\perp^4$ compared to the $\COcSa$
channel~\cite{Kramer:2001hh}. Therefore, although suppressed by
powers of $v$, CO contributions are important for $\jpsi$
production, especially at large $p_\perp$. We refer interested
readers to ref.~\cite{Ma:2010jj} for further discussion.},
\begin{align}\label{eq:states}
\mops, \quad \mopa, \quad\mopb, \quad\mopjz \, .
\end{align}
There are two other $P$-wave CO LDMEs that contribute to $J/\psi$
production with the same power counting as the $\mopjz$. However,
one can use heavy quark spin symmetry to relate $P$-wave operators
with $J=1, 2$ to the operator with $J=0$~\cite{Bodwin:1994jh},
\begin{align}
\mopj=(2J+1)\mopjz \left[1+O(v^2)\right].
\end{align}
For completeness, we list essential heavy quark pair states for
common heavy quarkonia production in table \ref{table}.
\begin{table}
\centering
\begin{tabular}{| c | c |}
  \hline
  Quarkonium & contributing states \\
  \hline
  $J/\psi$, $\psi'$, $\Upsilon(nS)$  & $^3S_1^{[1]}$, $^1S_0^{[8]}$,  $^3S_1^{[8]}$,   $^3P_J^{[8]}$ \\
  $\eta_{c}$, $\eta_b$ & $^1S_0^{[1]}$  \\
  $h_c$, $h_b$  & $^1P_1^{[1]}$,  $^1S_0^{[8]}$  \\
  $\chi_{cJ}$, $\chi_{bJ}$ &  $^3P_J^{[1]}$,  $^3S_1^{[8]}$ \\
  \hline
\end{tabular}
\caption{Essential heavy quark pair states for quarkonium
production. The contribution of color singlet states for each
quarkonium production is at leading power in $v$. The color octet
contributions for $P$-wave quarkonium production, say $h_{c,b}$ and
$\chi_{cJ, bJ}$, are also at leading power in $v$. The color octet
contributions to $S$-wave quarkonium production are power
suppressed.} \label{table}
\end{table}

The CGC enters the quarkonium framework in the derivation of the
perturbative cross-section $d \hat{\sigma}^\kappa$. We begin with
the heavy quark pair production amplitude in eq.~\eqref{eq:QQamp}
and project it on to a definite quantum configuration $\kappa$
\cite{Cho:1995vh} of the produced intermediate heavy quark pair,
\begin{align}\label{eq:amp0}
\begin{split}
M^{\kappa, J_z, (1, 8c)}(p)=&\sqrt{\frac{1}{m}}\sum_{L_z,
S_z}\sum_{s, \bar{s}}\sum_{i, \bar{i}} \left\langle L L_z; S S_z|J
J_z\right\rangle \left\langle\frac{1}{2}s; \frac{1}{2}\bar{s}|S
S_z\right\rangle
\left\langle3i;\bar{3}\bar{i}|(1,8c)\right\rangle\\
& \times
\begin{cases}
M^F_{s \bar{s}; i \bar{i}}(p,0),\quad \text{if $\kappa$ is $S$-wave},\\
\epsilon^{*}_\beta(L_z) M^{F, \beta}_{s \bar{s}; i
\bar{i}}(p,0),\quad \text{if $\kappa$ is $P$-wave},
\end{cases}
\end{split}
\end{align}
where $\epsilon^{*}_\beta(L_z)$ are polarization vectors discussed
further in appendix~\ref{sec:projectors}, and $M^{F, \beta}_{s
\bar{s}; i \bar{i}}(p,0)=\left. \frac{\partial}{\partial q^{\beta}}
M^{F}_{s \bar{s}; i \bar{i}}(p,q)\right|_{q=0}$. $(1, 8c)$ gives $1$
if $\kappa$ is CS, and $8c$ if $\kappa$ is CO. The color and spin
quantum numbers for the heavy quark pair are projected out by the
sums over the respective $SU(3)$ and $SU(2)$ color and spin
Clebsch-Gordan coefficients
$\left\langle3i;\bar{3}\bar{i}|1\right\rangle=\delta_{i\bar{i}}/\sqrt{N_c}$,
$\left\langle3i;\bar{3}\bar{i}|8c\right\rangle=\sqrt{2}t^{c}_{i\bar{i}}$
and $\left\langle\frac{1}{2}s; \frac{1}{2}\bar{s}|S
S_z\right\rangle$. The coefficients $\left\langle L L_z; S S_z|J
J_z\right\rangle$ account for the spin-orbit $LS$ coupling. As we
normalize the Dirac spinors as $\bar{u}u=-\bar{v}v=2m$, and
normalize the heavy quark pair composite state as $\left\langle
Q\bar{Q}(\kappa)|Q\bar{Q}(\kappa)\right\rangle =4m$, we have the
extra normalization factor
$\sqrt{\frac{1}{m}}=\frac{\sqrt{4m}}{\sqrt{2m}\sqrt{2m}}$.

To simplify our notation, we will suppress the color index in the rest
of the paper by introducing the matrix notation
\begin{align}\label{color}
\mathcal{C}^\kappa=
\begin{cases}
\mathcal{C}^{[1]}=\frac{{\bf 1}}{\sqrt{N_c}},\quad \text{if $\kappa$ is CS},\\
\mathcal{C}^{[8]}=\sqrt{2}t^{c},\quad \text{if $\kappa$ is CO}\,,
\end{cases}
\end{align}
where ${\bf 1}$ is a unit $3\times 3$ matrix. Then distinguishing
the color structure from the spinor structure, we can rewrite
eq.~\eqref{eq:amp0} as
\begin{align}\label{eq:amp}
\begin{split}
&M^{\kappa, J_z}(p)=\frac{g_s^2}{(2\pi)^4} \underset{\vka,\vk}{\int}
\frac{\rho_{p,a}(x_p,\vka)}{k_{1\perp}^2} \underset{\vx, \vy}\int
e^{i
\vk\cdot\vx} e^{i (\vp-\vk-\vka)\cdot\vy}\\
&\times \left\{ \text{Tr}\left[\mathcal{C}^{\kappa} V_F(\vx)t^a
V_F^\dagger(\vy)\right] \mathcal{F}^{\kappa,
J_z}_{q\bar{q}}\left(p,\vka,\vk\right) +
\text{Tr}\left[\mathcal{C}^{\kappa} t^b V_A^{b a}(\vx)\right]
\mathcal{F}^{\kappa, J_z}_g(p,\vka)\right\}\,,
\end{split}
\end{align}
where the functions $\mathcal{F}^{\kappa,
J_z}_{q\bar{q}}\left(p,\vka,\vk\right)$ and $\mathcal{F}^{\kappa,
J_z}_g(p,\vka)$ are defined as
\begin{subequations}\label{Tqqg}
\begin{align}
\begin{split}
\mathcal{F}^{\kappa,
J_z}_{q\bar{q}}\left(p,\vka,\vk\right)=&\sum_{L_z,
S_z}\left\langle L L_z; S S_z|J J_z\right\rangle\\
&\hspace{-1cm}\times\begin{cases} \left. \text{Tr}\left[\Pi^{SS_z}
T_{q\bar{q}}\left(p,q,\vka,\vk\right)\right]\right|_{q=0},\quad \text{if $\kappa$ is $S$-wave},\\
\epsilon^{*}_\beta(L_z) \left. \frac{\partial}{\partial q^{\beta}}
\text{Tr}\left[\Pi^{SS_z}
T_{q\bar{q}}\left(p,q,\vka,\vk\right)\right]\right|_{q=0},\quad
\text{if $\kappa$ is $P$-wave},
\end{cases}
\end{split}
\\
\begin{split}
\mathcal{F}^{\kappa, J_z}_g\left(p,\vka\right)=&\sum_{L_z,
S_z}\left\langle L L_z; S S_z|J J_z\right\rangle\\
&\times\begin{cases} \left. \text{Tr}\left[\Pi^{SS_z}
T_g\left(p,\vka\right)\right]\right|_{q=0},\quad \text{if $\kappa$ is $S$-wave},\\
\epsilon^{*}_\beta(L_z) \left. \frac{\partial}{\partial q^{\beta}}
\text{Tr}\left[\Pi^{SS_z}
T_g\left(p,\vka\right)\right]\right|_{q=0},\quad \text{if $\kappa$
is $P$-wave},
\end{cases}
\end{split}
\end{align}
\end{subequations}
with covariant spin projectors given
by~\cite{Kuhn:1979bb,Guberina:1980dc}
\begin{align}
\Pi^{S S_z}=\sqrt{\frac{1}{m}}\sum_{s,
\bar{s}}\left\langle\frac{1}{2}s; \frac{1}{2}\bar{s}|S
S_z\right\rangle v_{\bar{s}}(\frac{p}{2}-q)
\bar{u}_{{s}}(\frac{p}{2}+q),
\end{align}
with
\begin{subequations}\label{spin}
\begin{align}
\Pi^{00}=&\frac{1}{\sqrt{8m^3}}\left(\frac{\slashed{p}}{2}-\slashed{q}-m\right)
\gamma^5\left(\frac{\slashed{p}}{2}+\slashed{q}+m\right),\\
\Pi^{1S_z}=&\frac{1}{\sqrt{8m^3}}\left(\frac{\slashed{p}}{2}-\slashed{q}-m\right)
\slashed{\epsilon}^*(S_z)\left(\frac{\slashed{p}}{2}+\slashed{q}+m\right).
\end{align}
\end{subequations}

After these color and spin projections, the probability
$P^\kappa_1(\vb)$ to produce a heavy quark pair at an impact
parameter $\vb$ can be obtained as follows. One first squares the
spin and color projected amplitude. Next, averages are performed
over all possible color charge densities in both proton and nucleus.
Finally, the degrees of freedom of the heavy quark pair with quantum
number $\kappa$ are averaged over\footnote{To understand why one
averages over the states of the heavy quark pair, let us go back to
the NRQCD factorization formula in eq.~\eqref{eq:factorization}.
Assume that there are $N^\kappa$ possible states for each
configuration $\kappa$. We can denote these by $\lambda_1$,
$\cdots$, $\lambda_{N^\kappa}$. Then the factorization formula is
\begin{align}
d\sigma_H=\sum_\kappa
\sum_{\lambda_\kappa=1,\cdots,\lambda_{N^\kappa}} d
\hat{\sigma}^{\kappa,\lambda_\kappa}
\langle\mo^{H}_{\kappa,\lambda_\kappa}\rangle.\nonumber
\end{align}
Heavy quark spin symmetry as well as rotational invariance in color
space imply that the matrix elements
$\langle\mo^{H}_{\kappa,\lambda_\kappa}\rangle$ are independent of
$\lambda_\kappa$. If we then define the LDMEs as the summation of
all possible states, $\langle\mo^{H}_\kappa\rangle=
\sum_{\lambda_\kappa=1,\cdots,\lambda_{N^\kappa}}
\langle\mo^{H}_{\kappa,\lambda_\kappa}\rangle$, the NRQCD
factorization formula in eq.~\eqref{eq:factorization} is defined to
be $d
\hat{\sigma}^{\kappa}=\frac{1}{N^\kappa}\sum_{\lambda_\kappa=1,\cdots,\lambda_{N^\kappa}}
d \hat{\sigma}^{\kappa,\lambda_\kappa}$.}.

For the complex conjugate amplitude, we will denote all Lorentz, color and spin indices, as well as unobserved
momenta and coordinates, by a prime in their top right corner. Thus
$P^\kappa_1(\vb)$ can be written as
\begin{align}\label{eq:P1l}
\begin{split}
&P^\kappa_1(\vb)=\int
\left[\mathcal{D}\rho_p\right]\left[\mathcal{D}\rho_A\right]
W_p(x_p,\rho_p)W_A(x_A,\rho_A)
\frac{1}{N^\kappa}\sum_{\text{color}}\sum_{J_z}\left|M^{\kappa,
J_z}(p)\right|^2
\frac{d^3p}{(2\pi)^32E}\\
= \;&\frac{g_s^4}{(2\pi)^8} \int \frac{d^3p}{(2\pi)^32E}
\underset{\vka,\vk, \vkap,\vkp}{\int}
\frac{\left\langle\rho_{p,a}(x_p,\vka)\rho_{p,a'}^\dagger(x_p,\vkap)\right\rangle_{y_p}}{k_{1\perp}^2
k_{1\perp}^{'2}}\\
& \times \underset{\vx, \vy,\vxp,\vyp}\int e^{i\left[ \vk\cdot\vx-
\vkp\cdot\vxp+ (\vp-\vk-\vka)\cdot\vy- (\vp-\vkp-\vkap)\cdot\vyp\right]}\\
&\times \frac{1}{N^\kappa}\sum_{J_z}\left\{
\left\langle\text{Tr}\left[\mathcal{C}^\kappa V_F(\vx)t^a
V_F^\dagger(\vy)\right]
\text{Tr}\left[V_F(\vyp)t^{a'}V_F^\dagger(\vxp)\mathcal{\hat
C}^\kappa \right]\right\rangle_{y_A}\right.\\
&\hspace{2.5cm}\times\mathcal{F}^{\kappa,
J_z}_{q\bar{q}}\left(p,\vka,\vk\right)
\mathcal{F}^{\dagger\kappa, J_z}_{q\bar{q}}\left(p,\vkap,\vkp\right)\\
&~~~+\left\langle\text{Tr}\left[\mathcal{C}^\kappa V_F(\vx)t^a
V_F^\dagger(\vy)\right]\text{Tr}\left[V_A^{\dagger
a'b'}(\vxp)t^{b'}\mathcal{C}^\kappa \right]\right\rangle_{y_A}
\mathcal{F}^{\kappa, J_z}_{q\bar{q}}\left(p,\vka,\vk\right)\mathcal{F}^{\dagger\kappa, J_z}_g(p,\vkap)\\
&~~~+\left\langle\text{Tr}\left[\mathcal{C}^\kappa t^b V_A^{b
a}(\vx)\right]
\text{Tr}\left[V_F(\vyp)t^{a'}V_F^\dagger(\vxp)\mathcal{\hat
C}^\kappa \right]\right\rangle_{y_A} \mathcal{F}^{\kappa,
J_z}_g(p,\vka)
\mathcal{F}^{\dagger\kappa, J_z}_{q\bar{q}}\left(p,\vkap,\vkp\right)\\
&~~~+\left.\left\langle\text{Tr}\left[\mathcal{C}^\kappa t^b V_A^{b
a}(\vx)\right]\text{Tr}\left[V_A^{\dagger
a'b'}(\vxp)t^{b'}\mathcal{C}^\kappa \right]\right\rangle_{y_A}
\mathcal{F}^{\kappa, J_z}_g(p,\vka) \mathcal{F}^{\dagger\kappa,
J_z}_g(p,\vkap)\right\}\, .
\end{split}
\end{align}
Here $y_p=\ln(1/x_p)$ is the rapidity of the gluon that comes from
the proton, and $y_A=\ln(1/x_A)$ is the rapidity at which the Wilson
line correlators of the target nucleus are evaluated. In this
expression, $\langle \cdots \rangle_{y_{p(A)}}$ denotes the average
over color charge densities
\begin{align}
\langle {\cal O} [\rho_{p(A)}] \rangle_{y_{p(A)}} =
\int[d\rho_{p(A)}] W_{p(A)}(x_{p(A)},\rho_{p(A)}) {\cal O}
[\rho_{p(A)}] \label{eq:CGC-operator}\;,
\end{align}
where ${\cal O}$ here generically denotes the average over the
projectile charge density $\rho_p$ or the target color charge
density $\rho_A$ in eq.~(\ref{eq:P1l}). Further, the summation over
color degrees of freedom after the second equal sign has been taken
care of by our default rule: any repeated indices are assumed to be
summed over. $N^\kappa=(2J+1) N^{\text{color}}$ are the number of
states for a given $\kappa$, with $N^{\text{color}}=1~\text{or}~
N_c^2-1$ if $\kappa$ is color singlet or color octet, respectively.
For convenience, we will use
\begin{align}
\overline{\sum_{J_z}}&\equiv\frac{1}{2J+1}\sum_{J_z},\\
\overline{\mathcal{C}}^\kappa&\equiv\frac{\mathcal{
C}^\kappa}{\sqrt{N^{\text{color}}}},
\end{align}
in the rest of the paper.

All transverse coordinates in eq.~\eqref{eq:P1l} are defined with
respect to the center of the proton. To convert these to the
coordinates with respect to the center of nucleus, one simply has to
shift all coordinates
 by the impact parameter $\vb$. (For example,  $\vx \to \vx
-\vb$.)  Translational invariance guarantees\footnote {This assumes
that the size of nucleus is large enough for translational
invariance to apply.} that the averaged values in $\langle\cdots
\rangle_{y_A}$ are unchanged under such a shift. Therefore such a
shift only leads to the extra phase factor $e^{i
(\vka-\vkap)\cdot\vb}$.

When we derive the cross section $d\hat{\sigma}^\kappa$ for a  minimum bias proton-nucleus
collision, we have to integrate $P^\kappa_1(\vb)$ over
the impact parameter $\vb$. This generates the factor
\begin{align}
\underset{\vb}{\int} e^{i (\vka-\vkap)\cdot\vb}
=(2\pi)^2\delta^2(\vka-\vkap)\,  .
\end{align}
Using the delta function to integrate out the $\vkap$, we find that
the average over color density on the proton side to be
$\left\langle\rho_{p,a}(x_p,\vka)\rho_{p,a'}^\dagger(x_p,\vka)\right\rangle_{y_p}$.
Following \cite{Blaizot:2004wv}, we define the unintegrated gluon
distribution inside the proton to be\footnote{The unintegrated gluon
distribution in eq.~(\ref{eq:corr-proton}) is normalized such that
the leading log gluon distribution in the proton satisfies $$ x_p
f_{p/g}(x_p,Q^2)= {1\over 4\pi^3}\, \int_0^{Q^2} d l_\perp^2
\varphi_p(x_p,l_\perp)\; .$$See eq.~(\ref{eq:collinear-fact}) and
ref.~\cite{Fujii:2006ab} for further discussion.}
\begin{align}\label{eq:corr-proton}
g_s^2
\left\langle\rho_{p,a}(x_p,\vka)\rho_{p,a'}^\dagger(x_p,\vka)\right\rangle_{y_p}
=\frac{\delta_{a a'}}{\pi(N_c^2-1)} k_{1\perp}^2
\varphi_{p,y_p}(\vka).
\end{align}

With this substitution,  the differential cross section of
production of heavy quark pair with quantum number $\kappa$ can be
written as
\begin{align}\label{eq:ds0}
\begin{split}
&\frac{d \hat{\sigma}^\kappa}{d^2\vp d y}=\frac{g_s^2}{(2\pi)^{10}
(N_c^2-1)} \underset{\vka,\vk, \vkp}{\int}
\frac{\varphi_{p,y_p}(\vka)}{k_{1\perp}^2}\\
&\times \underset{\vx, \vy,\vxp,\vyp}\int e^{i\left[ \vk\cdot\vx-
\vkp\cdot\vxp+ (\vp-\vk-\vka)\cdot\vy- (\vp-\vkp-\vka)\cdot\vyp\right]}\\
&\times \overline{\sum_{J_z}}\left\{
\left\langle\text{Tr}\left[\overline{\mathcal{C}}^\kappa V_F(\vx)t^a
V_F^\dagger(\vy)\right]
\text{Tr}\left[V_F(\vyp)t^{a}V_F^\dagger(\vxp)\overline{\mathcal{C}}^\kappa
\right]\right\rangle_{y_A} \right.\\
&\hspace{2.5cm}\times\mathcal{F}^{\kappa,
J_z}_{q\bar{q}}\left(p,\vka,\vk\right)
\mathcal{F}^{\dagger\kappa, J_z}_{q\bar{q}}\left(p,\vka,\vkp\right)\\
&~~~+\left\langle\text{Tr}\left[\overline{\mathcal{C}}^\kappa
V_F(\vx)t^a V_F^\dagger(\vy)\right]\text{Tr}\left[V_A^{\dagger
ab'}(\vxp)t^{b'}\overline{\mathcal{C}}^\kappa
\right]\right\rangle_{y_A}
\mathcal{F}^{\kappa, J_z}_{q\bar{q}}\left(p,\vka,\vk\right)\mathcal{F}^{\dagger\kappa, J_z}_g(p,\vka)\\
&~~~+\left\langle\text{Tr}\left[\overline{\mathcal{C}}^\kappa t^b
V_A^{b a}(\vx)\right]
\text{Tr}\left[V_F(\vyp)t^{a}V_F^\dagger(\vxp)\overline{\mathcal{C}}^\kappa
\right]\right\rangle_{y_A} \mathcal{F}^{\kappa, J_z}_g(p,\vka)
\mathcal{F}^{\dagger\kappa, J_z}_{q\bar{q}}\left(p,\vka,\vkp\right)\\
&~~~+\left.\left\langle\text{Tr}\left[\overline{\mathcal{C}}^\kappa
t^b V_A^{b a}(\vx)\right]\text{Tr}\left[V_A^{\dagger
ab'}(\vxp)t^{b'}\overline{\mathcal{C}}^\kappa
\right]\right\rangle_{y_A} \mathcal{F}^{\kappa, J_z}_g(p,\vka)
\mathcal{F}^{\dagger\kappa, J_z}_g(p,\vka)\right\}.
\end{split}
\end{align}
This expression is the main result of our paper. With the NRQCD
color projectors in eq.~\eqref{color}, we can work out all the
Wilson lines expectation values in the above equation. The
expression for $\mathcal{F}^{\kappa, J_z}_{q\bar{q}},
\mathcal{F}^{\kappa, J_z}_g$ in eq.~\eqref{Tqqg} along with the
NRQCD spin projectors in eq.~\eqref{spin}, allow us to derive all
the partonic hard-part functions. In the rest of the paper, we will
work out the explicit simplifications of this general result for the
color singlet and color octet channels in the large $N_c$ limit. The
phenomenological applications of this result will be left for future
publications.

\subsection{Complete results for quarkonium cross-sections in the large $N_c$ limit}\label{sec:color}

\subsubsection{Color singlet contributions}

If $\kappa$ is a color singlet intermediate state, only the first
term $\propto \mathcal{F}^{\kappa, J_z}_{q\bar
q}\mathcal{F}^{\dagger\kappa, J_z}_{q\bar q}$ in eq.~\eqref{eq:ds0}
survives;  all other terms vanish. This is because all other terms
involve $\mathcal{F}^{\kappa, J_z}_g$ (or
$\mathcal{F}^{\dagger\kappa, J_z}_g$), in which a gluon naturally
transforms into a color octet heavy quark pair state. Taking
$\overline{\mathcal{C}}^{\kappa}=\overline{\mathcal{C}}^{[1]}=\frac{\bm{1}}{\sqrt{N_c}}$,
we find
\begin{align}\label{singlet-1}
\begin{split}
&\left\langle\text{Tr}\left[\overline{\mathcal{C}}^{[1]} V_F(\vx)t^a
V_F^\dagger(\vy)\right]
\text{Tr}\left[V_F(\vyp)t^{a}V_F^\dagger(\vxp)\overline{\mathcal{C}}^{[1]}
\right]\right\rangle_{y_A}\\
=& \frac{1}{2}\left\{Q_{\vx\vxp\vyp\vy}^{y_A}-\frac{1}{N_c^2}\left\langle
\text{Tr}\left[V_F(\vx) V_F^\dagger(\vy)\right]
\text{Tr}\left[V_F(\vyp)V_F^\dagger(\vxp)\right]
\right\rangle_{y_A}\right\},
\end{split}
\end{align}
where we have used the identity
\begin{align}\label{eq:Fierz}
\sum_a t^a_{ij}t^a_{kl}=\frac{1}{2}\left(\delta_{il}\delta_{kj}
-\frac{1}{N_c} \delta_{ij}\delta_{kl}\right)\, .
\end{align}
In eq.~(\ref{singlet-1}), $Q_{\vx,\vxp,\vyp,\vy}$ is the quadrupole
correlator
\begin{align}
Q_{\vx,\vxp,\vyp,\vy}^{y_A} \equiv \frac{1}{N_c}\left\langle
\text{Tr}\left[V_F(\vx)
V_F^\dagger(\vxp)V_F(\vyp)V_F^\dagger(\vy)\right]\right\rangle_{y_A}.
\end{align}
Further, in the large $N_c$ limit and for large nuclei ($\alpha_s^2
A^{1/3} >>1$), the expectation value for the second term in
eq.~\eqref{singlet-1} can be factored as the product of the
expectation values of the traces within as
\begin{align}
\begin{split}
&\left\langle \text{Tr}\left[V_F(\vx) V_F^\dagger(\vy)\right]
\text{Tr}\left[V_F(\vyp)V_F^\dagger(\vxp)\right]\right\rangle_{y_A}\\
\to &\left\langle \text{Tr}\left[V_F(\vx)
V_F^\dagger(\vy)\right]\right\rangle_{y_A} \left\langle
\text{Tr}\left[V_F(\vyp)V_F^\dagger(\vxp)\right]\right\rangle_{y_A}.
\end{split}
\end{align}
Using translation invariance for large nuclei, one can express the well known dipole correlator as
\begin{align}
D_{\vx-\vxp}^{y_A}=D_{\vxp-\vx}\equiv\frac{1}{N_c}\left\langle
\text{Tr}\left[V_F(\vx) V_F^\dagger(\vxp)\right]\right\rangle_{y_A}
.
\end{align}

Thus in the large $N_c$ and large $A$ limit, the expectation value
over color charge densities in the nucleus in eq.~(\ref{singlet-1})
can be expressed as
\begin{align}\label{eq:CScolor}
\begin{split}
&\left\langle\text{Tr}\left[\overline{\mathcal{C}}^{[1]} V_F(\vx)t^a
V_F^\dagger(\vy)\right]
\text{Tr}\left[V_F(\vyp)t^{a}V_F^\dagger(\vxp)\overline{\mathcal{C}}^{[1]}
\right]\right\rangle_{y_A}\\
=& \frac{1}{2} \left(Q_{\vx,\vxp,\vyp,\vy}^{y_A} -D_{\vx-\vy}^{y_A}
D_{\vyp-\vxp}^{y_A} \right).
\end{split}
\end{align}
Henceforth, for simplicity of notation, we will not write out explicitly the rapidity index on the quadrupole and dipole correlators.

It is convenient to express our result in terms of the variables
$\vrz$, $\vD$, $\vr$, and $\vrp$ which are expressed in terms of the
co-ordinates $\vx$, $\vxp$, $\vyp$, and $\vy$ as
\begin{align}
\begin{split}
\vx=&\vrz+\frac{\vr}{2}, \quad\quad\quad\quad \vy=\vrz-\frac{\vr}{2},\\
\vxp=&\vD+\vrz+\frac{\vrp}{2}, \quad \vyp=\vD+\vrz-\frac{\vrp}{2}.
\end{split}
\end{align}
Translation invariance implies that eq.~\eqref{eq:CScolor} is
independent of $\vrz$. The $\vrz$ integration can therefore be
performed trivially, giving as a result $\pi R_A^2$, the transverse
area of the nucleus.

With these coordinate transformations, we obtain the cross-section
for the production of color singlet heavy quark pairs to be
\begin{align}\label{eq:dsktCS}
\begin{split}
\frac{d \hat{\sigma}^\kappa}{d^2\vp d
y}\overset{\text{CS}}=&\frac{\alpha_s \pi R_A^2}{(2\pi)^{7}
(N_c^2-1)} \underset{\vka}{\int}
\frac{\varphi_{p,y_p}(\vka)}{k_{1\perp}^2}\underset{\vD,\vr,\vrp}\int
e^{-i(\vp-\vka)\cdot\vD} \\
&\times
\left(Q_{\frac{\vr}{2},\vD+\frac{\vrp}{2},
\vD-\frac{\vrp}{2},-\frac{\vr}{2}} -D_{\vr}
D_{\vrp} \right) \Gamma_1^\kappa,
\end{split}
\end{align}
where $\Gamma_1^\kappa$ are defined as
\begin{align}\label{eq:defgamma1}
\Gamma_1^\kappa \equiv \frac{1}{(2\pi)^2} \underset{\vk, \vkp}{\int}
e^{i (\vk-\frac{\vp-\vka}{2})\cdot\vr}e^{-i
(\vkp-\frac{\vp-\vka}{2})\cdot\vrp} \overline{\sum_{J_z}}
\mathcal{F}^{\kappa, J_z}_{q\bar{q}}\left(p,\vka,\vk\right)
\mathcal{F}^{\dagger\kappa, J_z}_{q\bar{q}}\left(p,\vka,\vkp\right),
\end{align}
which are listed in appendix~\ref{sec:cshard}. Note that, if
$\Gamma^\kappa_1\propto\delta(\vr)$ or $\delta(\vrp)$, the
quadrupole correlator in eq.~\eqref{eq:dsktCS} collapses to a single
dipole correlator and cancels the second term exactly. Thus the
terms in $\Gamma^\kappa_1$ that are proportional to $\delta(\vr)$ or
$\delta(\vrp)$ do not contribute to the heavy quarkonium cross
section and shall be neglected.

In the limit of $\nc\rightarrow \infty$ and $\alpha_s^2
A^{1/3}\rightarrow \infty$, the dipole correlator in
eq.~(\ref{eq:dsktCS}) satisfies the Balitsky-Kovchegov (BK)
equation~\cite{Balitsky:1995ub,Kovchegov:1999yj},
\begin{align}
\begin{split}
 {d\over dy_A} D (\xt - \yt) =&
{N_c\, \alpha_s \over 2\pi^2} \int\, d^2 \zt\, {(\xt - \yt)^2 \over
(\xt - \zt)^2 (\zt - \yt)^2}\;\\
&\times\left[D (\xt - \zt)\, D (\zt - \yt) - D (\xt - \yt)\right]\,
.
\end{split}
\label{eq:2pt}
\end{align}
In the low density limit $|\xt-\yt| Q_{s} <<1$, this equation
reduces to the well known BFKL
equation~\cite{Kuraev:1977fs,Balitsky:1978ic}, which describes the
leading logarithmic behavior of perturbative QCD at small $x$. The
BK equation is the simplest equation of high energy QCD describing
both small $x$ QCD evolution and coherent multiple scattering and is
used widely in phenomenological applications in both deeply
inelastic scattering and hadron-hadron scattering.

The quadrupole correlator in eq.~(\ref{eq:dsktCS}) is less well
known but is an equally fundamental object in high energy QCD.
Evolution equations in the JIMWLK framework for the quadrupole have
been derived~\cite{JalilianMarian:2004da}. Their evolution can be
computed numerically~\cite{Dumitru:2011vk} and analytic results
obtained in different limits~\cite{Iancu:2011nj}. It has been argued
that in the large $\nc$ limit, dipole and quadrupole operators are
the only universal multi-gluon correlators that appear in the
``dilute-dense" final states~\cite{Dominguez:2012ad}. This theorem
certainly appears to hold for quarkonium production in the color
singlet channel and, as we shall shortly discuss, in the color octet
channel.

\subsubsection{Color octet contributions}

For the color octet state $\kappa$,
$\overline{\mathcal{C}}^{\kappa}=\overline{\mathcal{C}}^{[8]}=\frac{\sqrt{2}t^c}{\sqrt{N_c^2-1}}$,
the first term in eq.~\eqref{eq:ds0} gives
\begin{align}\label{eq:part11}
\begin{split}
&\left\langle\text{Tr}\left[\overline{\mathcal{C}}^{[8]} V_F(\vx)t^a
V_F^\dagger(\vy)\right]
\text{Tr}\left[V_F(\vyp)t^{a}V_F^\dagger(\vxp)\overline{\mathcal{C}}^{[8]}
\right]\right\rangle_{y_A}\\
=& \frac{1}{2(N_c^2-1)}\left\langle \text{Tr}\left[V_F(\vx)
V_F^\dagger(\vxp)\right]
\text{Tr}\left[V_F(\vyp)V_F^\dagger(\vy)\right]\right.\\
& -\frac{1}{N_c} \text{Tr}\left[V_F(\vx)
V_F^\dagger(\vy)V_F(\vyp)V_F^\dagger(\vxp)\right]\\
&-\frac{1}{N_c} \text{Tr}\left[V_F(\vx)
V_F^\dagger(\vxp)V_F(\vyp)V_F^\dagger(\vy)\right]\\
&\left.  +\frac{1}{N_c^2} \text{Tr}\left[V_F(\vx)
V_F^\dagger(\vy)\right]
\text{Tr}\left[V_F(\vyp)V_F^\dagger(\vxp)\right] \right\rangle_{y_A}
\, .
\end{split}
\end{align}
Here we have used the identity in eq.~\eqref{eq:Fierz} repeatedly.
The expression in eq.~\eqref{eq:part11} can be significantly
simplified if we take the large $N_c$ limit. In this limit, the
first term in eq.~\eqref{eq:part11} dominates since it scales as
$O(N_c^2)$ while all the other terms scale as $O(1)$ in color space.
We thus obtain
\begin{align}
&\left\langle\text{Tr}\left[\overline{\mathcal{C}}^{[8]} V_F(\vx)t^a
V_F^\dagger(\vy)\right]
\text{Tr}\left[V_F(\vyp)t^{a}V_F^\dagger(\vxp)\overline{\mathcal{C}}^{[8]}
\right]\right\rangle_{y_A} \to\frac{1}{2} D_{\vx-\vxp}D_{\vy-\vyp}.
\end{align}

Defining the dipole unintegrated gluon distribution in momentum space as
\begin{align}
\mathcal{N}(\vk)=\mathcal{N}(-\vk)\equiv\underset{\vr}\int e^{i
\vk\cdot\vr} D_{\vr}.
\end{align}
 one can integrate out all the coordinate variables in eq.~\eqref{eq:ds0} straightforwardly, and obtain
\begin{align}
\begin{split}
&\underset{\vx, \vy,\vxp,\vyp}\int e^{i\left[ \vk\cdot\vx-
\vkp\cdot\vxp+ (\vp-\vk-\vka)\cdot\vy-
(\vp-\vkp-\vka)\cdot\vyp\right]} D_{\vx-\vxp}D_{\vy-\vyp}\\
=& (2\pi)^2\delta^2(\vk-\vkp)\pi R_A^2\;
\mathcal{N}(\vk)\;\mathcal{N}(\vp-\vka-\vk).
\end{split}
\end{align}
As a result, the first term in the braces in eq.~\eqref{eq:ds0}
gives
\begin{align}\label{eq:part12}
\begin{split}
\frac{g_s^2 (\pi R_A^2)}{2(2\pi)^{8} (N_c^2-1)}
&\underset{\vka,\vk}{\int}\!\!
\frac{\varphi_{p,y_p}(\vka)}{k_{1\perp}^2}\mathcal{N}(\vk)\;\mathcal{N}(\vp-\vka-\vk)\\
& \times \overline{\sum_{J_z}} \mathcal{F}^{\kappa,
J_z}_{q\bar{q}}\left(p,\vka,\vk\right) \mathcal{F}^{\dagger\kappa,
J_z}_{q\bar{q}}\left(p,\vka,\vk\right).
\end{split}
\end{align}

Likewise, we can work out the color algebra for the remaining three
terms in eq.~\eqref{eq:ds0} making use of the identity
\begin{align}
V_A^{ab}(\vr)=2\text{Tr}\left[V_F^\dagger(\vr)t^a V_F(\vr)t^{b}\,
\right].
\end{align}
Adding up all these terms together, we find
\begin{align}\label{eq:dsktCO}
\frac{d \hat{\sigma}^\kappa}{d^2\vp d
y}\overset{\text{CO}}=\frac{\alpha_s (\pi R_A^2)}{(2\pi)^{7}
(N_c^2-1)} \underset{\vka,\vk}{\int}
\frac{\varphi_{p,y_p}(\vka)}{k_{1\perp}^2}\mathcal{N}(\vk)\mathcal{N}(\vp-\vka-\vk)
\Gamma^\kappa_8,
\end{align}
with
\begin{align}\label{eq:defgamma8}
\Gamma^\kappa_8\equiv\overline{\sum_{J_z}}
\left|\mathcal{F}^{\kappa, J_z}_{q\bar{q}}\left(p,\vka,\vk\right)
+\mathcal{F}^{\kappa, J_z}_g(p,\vka)\right|^2.
\end{align}
With the spin projectors in eq.~\eqref{spin}, the calculations of
$\Gamma^\kappa_8$ are straightforward, and we list the results in
appendix~\ref{sec:cohard}. Note that unlike the case in the color
singlet channel, only dipole correlators appear in the color octet
channels.

Eqs.~\eqref{eq:dsktCS} and \eqref{eq:dsktCO} represent our complete
expressions for heavy quarkonium production under the large $N_c$
limit. The corresponding functions for the hard matrix elements
$\Gamma^\kappa_1$ and $\Gamma^\kappa_8$ are given in appendix
\ref{sec:cshard} and \ref{sec:cohard} for various heavy quark pair
states $^{2S+1}L_J^{[C]}$. Once these results are multiplied by the
corresponding NRQCD LDMEs $\langle\mo^{H}_\kappa\rangle$, one
obtains the differential cross-section for the production of heavy
quarkonium states in high energy proton-nucleus collisions. The
results collected in the appendix provide a complete set for
phenomenological studies of all the common heavy quarkonium states.

\subsection{The proton collinear limit}\label{sec:collinear}
When the gluon momentum fraction $x_p$ in the proton is not very small,
 the typical transverse momentum of the gluons in the proton is
much smaller than the mass and the transverse momentum of heavy
quarkonium, $Q_{s,p}(x_p)\ll k_{1\perp} \ll m$ and $Q_{s,p}(x_p)\ll
k_{1\perp} \ll p_\perp$. We can then take the limit $k_{1\perp}\to
0$ in both the hard part and in the Wilson lines. Then one can
integrate out $\vka$ and arrive at a collinear gluon distribution
function in the proton, thereby restoring collinear factorization
from the proton side. Using $d^2\vka=\frac{1}{2}d\theta_1
dk_{1\perp}^2$ and defining
\begin{align}
\frac{1}{4\pi^3}\int^{Q^2} \varphi_{p,y_p}(\vka) dk_{1\perp}^2
\equiv x_p f_{p/g}(x_p,Q^2)\,,
\label{eq:collinear-fact}
\end{align}
we find for the color singlet channel,
\begin{align}\label{eq:dscollCS}
\begin{split}
\frac{d \hat{\sigma}^\kappa}{d^2\vp d
y}\overset{\text{CS}}=&\frac{\alpha_s (\pi R_A^2)}{4(2\pi)^{3}
(N_c^2-1)} {x_p f_{p/g}(x_p,Q^2)}\underset{\vD,\vr,\vrp}\int
e^{-i\vp\cdot\vD}\\
&\times
\left(Q_{\left(\frac{\vr}{2}\right)\left(\vD+\frac{\vrp}{2}\right)
\left(\vD-\frac{\vrp}{2}\right)\left(-\frac{\vr}{2}\right)} -D_{\vr}
D_{\vrp} \right) \tilde{\Gamma}_1^\kappa\,,
\end{split}
\end{align}
where
\begin{align}\label{cs-limit}
\tilde{\Gamma}_1^\kappa \equiv \underset{k_{1\perp}\to
0}{\text{lim.}}\; \frac{1}{2\pi}\int_0^{2\pi} d\theta_1\;
\frac{\Gamma_1^\kappa}{k_{1\perp}^2},
\end{align}
which are listed in Appendix~\ref{sec:cshard}. Similarly, for the
color octet channel, we obtain
\begin{align}\label{eq:dscollCO}
\frac{d \hat{\sigma}^\kappa}{d^2\vp d
y}\overset{\text{CO}}=\frac{\alpha_s (\pi R_A^2)}{4(2\pi)^{3}
(N_c^2-1)} {x_p f_{p/g}(x_p,Q^2)} \underset{\vk}{\int}
\mathcal{N}(\vk)\;\mathcal{N}(\vp-\vk) \;\tilde{\Gamma}^\kappa_8\,,
\end{align}
with
\begin{align}\label{co-limit}
\tilde{\Gamma}^\kappa_8 \equiv \underset{k_{1\perp}\to
0}{\text{lim.}} \frac{1}{2\pi}\int_0^{2\pi} d\theta_1
\frac{\Gamma^\kappa_8}{k_{1\perp}^2}\, .
\end{align}
Detailed expressions  can be found in Appendix~\ref{sec:cohard}. It
is important to realize that both $\Gamma_1^\kappa$ and
$\Gamma_8^\kappa$ are quadratic in $\vka$ when $\vka\to 0$. Thus
$\tilde{\Gamma}^\kappa_1$ and $\tilde{\Gamma}^\kappa_8$ as defined
in eqs.~\eqref{cs-limit} and \eqref{co-limit} are both finite.

\subsection{Small $p_\perp$ limit}\label{sec:smallpt}

For simplicity, we will discuss the small $p_\perp$ behavior only in the proton collinear limit. Small $p_\perp$ behavior for general case can be obtained similarly. The kinematic regime we are considering here is $p_\perp \ll m$. Then eqs. \eqref{eq:dscollCS} and \eqref{eq:dscollCO} imply that the leading contribution region should be $k_\perp\sim k'_\perp\sim p_\perp \ll m$. We will derive the power law of $p_\perp/m$ for each channel. Variables $\tilde{X}_{l_\perp}$ and $\tilde{X}_{l'_\perp}$ have the behavior
\begin{align}
\tilde{X}_{l_\perp}=m^2+O(p_\perp^2),~~~~~~~~~~~~~~
\tilde{X}_{l'_\perp}=m^2+O(p_\perp^2).
\end{align}

For color singlet channels, we begin with eq.~\eqref{eq:gamma1coll}. It is easy to find
\begin{align}\label{eq:smallpt1}
\tilde{W}^{\CScPz} \sim \tilde{W}^{\CScPa} \sim \tilde{W}^{\CScPb} \sim \tilde{W}^{\CSaSz} \sim O\left(\frac{p_\perp^2}{m^2}\right).
\end{align}
Naively, it seems like $\tilde{W}^{\CScSa} \sim O(1)$ if we expand both $\tilde{X}_{l_\perp}$ and $\tilde{X}_{l'_\perp}$ to leading power. However, if $\tilde{X}_{l_\perp}$ is kept only to leading power, there is no $\vk$ dependence in $\tilde{W}^{\CScSa}$, which results in that $\tilde{\Gamma}_1^{\CScSa} \propto \delta^2(\vr)$. Substituting $\tilde{\Gamma}_1^{\CScSa} \propto \delta^2(\vr)$ into eq. \eqref{eq:dscollCS}, we find the expression vanishes. Therefore, to obtain a nonzero contribution, $\tilde{X}_{l_\perp}$ must be expanded to next-to-leading power. Similarly, $\tilde{X}_{l'_\perp}$ also needs to be expanded to next-to-leading power. Power law for $\tilde{W}^{\CSaPa}$ can be derived in the same way. We thus get
\begin{align}\label{eq:smallpt2}
\tilde{W}^{\CScSa} \sim \tilde{W}^{\CSaPa} \sim O\left(\frac{p_\perp^4}{m^4}\right).
\end{align}
For color octet channels, starting with eq.~\eqref{eq:gamma8coll}, and realizing
\begin{align}
1-\frac{m^2}{\tilde{X}_{l_\perp}} \sim O\left(\frac{p_\perp^2}{m^2}\right),
\end{align}
we can easily get
\begin{align}
&\tilde{\Gamma}_8^{\COcPz} \sim \tilde{\Gamma}_8^{\COcPb} \sim \tilde{\Gamma}_8^{\COcPj} \sim \tilde{\Gamma}_8^{\COaSz} \sim \tilde{\Gamma}_8^{\COaPa} \sim O\left(\frac{p_\perp^2}{m^2}\right)\label{eq:smallpt3},\\
&\tilde{\Gamma}_8^{\COcSa} \sim \tilde{\Gamma}_8^{\COcPa} \sim O\left(\frac{p_\perp^4}{m^4}\right)\label{eq:smallpt4}.
\end{align}

The power law of differential cross sections is complicated by different correlators between color singlet channel and color octet channel. If we assume the correlators do not contribute any power behaviors, then eqs. \eqref{eq:smallpt1}, \eqref{eq:smallpt2}, \eqref{eq:smallpt3} and \eqref{eq:smallpt4} are also the power law of differential cross section of each channel. We thus can discuss the relative importance of each channel. Taking $\jpsi$ production as an example, if we normalize the contribution of $\CScSa$ channel to be $O(1)$, then contribution of $\COcSa$ channel is $O(v^4)$, and contributions of $\COaSz$ channel and $\COcPj$ channel are $O(m^2v^4/p_\perp^2)$, where $v$ is the typical relative velocity between charm quark pair inside $\jpsi$. Therefore, the color singlet channel $\CScSa$ is dominant for $\jpsi$ production as long as $m \gg p_\perp \gg m v^2$. Conversely, for $p_\perp <  mv^2$, the color octet contribution will dominate. The latter regime was studied recently in ref.~\cite{Qiu:2013qka} --our results are in agreement with those presented there.

\subsection{Large $p_\perp$ limit}\label{sec:largept}

In the kinematic region $p_\perp\gg Q_{s}$, additional contributions
come from a higher order in $\alpha_s$ process where a recoiling
particle with large transverse momentum in the final state is needed
to balance the quarkonium's
$p_\perp$~\cite{Ferreiro:2008wc,Ferreiro:2009ur}. Nevertheless, we
can still study the limit $p_\perp \sim Q_s\gg m$, because $Q_{s}$
can be larger than $m$. To expand the hard matrix element in powers
of $m$ in this limit, we need to know the relative size of typical
values of $k_{1\perp}$.

Let us first consider the case where $p_\perp \sim Q_s \gg m \gg
k_{1\perp}$. In this case, all the results obtained in the previous
subsection (where we took the
 collinear limit for the proton side) are still valid. Normalizing the
$\COcSa$ channel as $O(1)$, from eqs.~\eqref{eq:gamma1coll} and
\eqref{eq:gamma8coll} we find that the $\CScSa$ channel behaves as
$m^4/p_\perp^4$.  All the other channels behave as $m^2/p_\perp^2$
if we restrict ourselves to the regime where $p_\perp \sim
l_\perp\sim l'_\perp$. The inclusion of
 other kinematic regions gives logarithm enhancements for
some channels; however, the power laws governing the $p_\perp$ dependence are not
changed.

Thus we find that at the perturbative order in our work color octet
channels will dominate large $p_\perp$ quarkonium production. This
is similar to the LO calculation for quarkonium production in
proton-proton collision using collinear
factorization~\cite{Kramer:2001hh}. In particular, for
$J^{PC}=1^{--}$ quarkonia such as $\jpsi$, contributions from the
color singlet channel are suppressed by $m^4/p_\perp^4$, implying
that color octet contributions may be large even if $p_\perp$ is not
too large.

From eqs.~\eqref{eq:gamma1} and \eqref{eq:gamma8}, we find that the
above power counting is unchanged if the typical value of
$k_{1\perp}$ is a little larger, $p_\perp\sim Q_s \gg m \sim
k_{1\perp}$. However, in the regime $p_\perp\sim Q_s \sim k_{1\perp}
\gg m$, although the $p_\perp$ power counting of
 of all other channels is unchanged, that for the
$\CScSa$ channel changes from $m^4/p_\perp^4$ to $m^2/p_\perp^2$.
The reason is that the contribution of the $\CScSa$ channel is
proportional to $k_{1\perp}^2 + 4m^2$. This can be seen in
eq.~\eqref{eq:gamma3s11}.

\section{Comparison with other approaches}\label{sec:comparison}
In this section, we discuss the relation between our complete NRQCD results and those from related theoretical works in the literature. In particular, we compare our results for the color singlet channel with those based on a quasi-classical saturation approach~\cite{Kharzeev:2005zr,Kharzeev:2008nw,Kharzeev:2008cv,Dominguez:2011cy,Kharzeev:2012py} and to results matching the CGC computations of \cite{Blaizot:2004wv} to the color evaporation model~\cite{Fujii:2006ab,Fujii:2013gxa}.

\subsection{Quasi-classical saturation model}\label{sec:CSM}
Within the framework of a quasi-classical approximation to the QCD
dipole model~\cite{Mueller:1993rr,Mueller:1994jq,Mueller:1994gb},
Dominguez et. al. investigated cold nuclear matter effects of
$\jpsi$ production in pA collisions in a series of
papers~\cite{Dominguez:2011cy,Kharzeev:2005zr,Kharzeev:2008nw,Kharzeev:2008cv,Kharzeev:2012py}.
Within the NRQCD factorization formalism, we naturally have both
color singlet and color octet contributions. We will compare here
our color singlet contribution with recent results in
\cite{Dominguez:2011cy,Kharzeev:2012py}.

Since the works of \cite{Dominguez:2011cy,Kharzeev:2012py} are
performed in the limit of collinear factorization on the proton
side, we will compare their results to our results for the color
singlet channel in collinear limit of eq.~\eqref{eq:dscollCS}. In
the quasi-classical approximation, the color sources in the nucleus
are assumed to be the Gaussian distributed sources of the
McLerran-Venugopalan model. As noted previously, this is a
Glauber-like multiple scattering
approximation~\cite{Blaizot:2004wu}.  In this quasi-classical
approximation, the quadrupole correlator in the large $\nc$ limit
reads~\cite{Blaizot:2004wv,Dominguez:2011wm}
\begin{align}
\begin{split}
Q_{\vx \vxp \vyp \vy}\approx & D_{\vx-\vy}D_{\vxp-\vyp}-\frac{\ln
(D_{\vx-\vyp}D_{\vxp-\vy}) - \ln(D_{\vx-\vxp}D_{\vy-\vyp})}{\ln
(D_{\vx-\vy}D_{\vxp-\vyp}) - \ln(D_{\vx-\vxp}D_{\vy-\vyp})}\\
&\times \left( D_{\vx-\vy}D_{\vxp-\vyp} - D_{\vx-\vxp}D_{\vy-\vyp}
\right).
\end{split}
\end{align}
Using the expression for $\tilde{\Gamma}_1^{\CScSa}$ in
appendix~\ref{sec:cshard} and the expression for the matrix element
in eq.~\eqref{eq:O3s11}, $\jpsi$ production in the color singlet
model gives
\begin{align}
\begin{split}
\frac{d {\sigma}^{\jpsi}}{d^2\vp d
y}\overset{\text{CSM}}=&\;\;\frac{3|R(0)|^2}{4\pi} \frac{d
\hat{\sigma}^{\CScSa}}{d^2\vp d y} \\
=&\frac{\alpha_s \pi R_A^2 m |R(0)|^2}{4(2\pi)^{4} N_c^2} {x_p
f_{p/g}(x_p,Q^2)}\underset{\vD,\vr,\vrp}\int
e^{-i\vp\cdot\vD} K_0(r_\perp m)K_0(r'_\perp m)\\
&\times
\frac{\ln\left[D_{\frac{1}{2}(\vr+\vrp)-\vD}D_{\frac{1}{2}(\vr+\vrp)+\vD}\right]-
\ln\left[D_{\frac{1}{2}(\vr-\vrp)-\vD}D_{\frac{1}{2}(\vr-\vrp)+\vD}\right]}{\ln
(D_{\vr}D_{\vrp}) -
\ln\left[D_{\frac{1}{2}(\vr-\vrp)-\vD}D_{\frac{1}{2}(\vr-\vrp)+\vD}\right]}\\
&\times \left[
D_{\frac{1}{2}(\vr-\vrp)-\vD}D_{\frac{1}{2}(\vr-\vrp)+\vD} -
D_{\vr}D_{\vrp}\right].
\end{split}
\end{align}
If we further change the integration variable $\vD\to-\vD$ and choose a Gaussian distribution for
the dipole correlator
\begin{align}\label{eq:mvapp}
D_{\vr}=e^{-\frac{1}{8}Q_{s}^2 r_\perp^2},
\end{align}
we arrive at a much simpler expression
\begin{align}\label{eq:cscomp}
\begin{split}
\frac{d {\sigma}^{\jpsi}}{d^2\vp d y}\overset{\text{CSM}}=&(\pi
R_A^2) x_p f_{p/g}(x_p,Q^2) \underset{\vD,\vr,\vrp}\int
\frac{e^{i\vp\cdot\vD}}{4(2\pi)^{4}} \Phi(r_\perp)\Phi(r'_\perp)\\
&\times \frac{4\vr\cdot\vrp}{(\vr+\vrp)^2-4\Delta_\perp^2}\left\{
e^{-\frac{Q_{s}^2}{16}[(\vr-\vrp)^2+4\Delta_\perp^2]}-
e^{-\frac{Q_{s}^2}{8}(r_\perp^2+{r'_\perp}^{2})}\right\},
\end{split}
\end{align}
where the wave-function $\Phi(r_\perp)$ is given by
\begin{align}
\Phi(r_\perp)\equiv \frac{g_s}{\pi\sqrt{2N_c}}\left[ m^2 K_0(r_\perp
m)\frac{ |R(0)|\sqrt{\pi}}{\sqrt{m^3} \sqrt{2N_c}}\right].
\end{align}

Remarkably, the above differential cross section is
equivalent\footnote{A careful reader will observe that the term
$4\vr\cdot\vrp$ in eq.~\eqref{eq:cscomp} is a little different from
the corresponding term in \cite{Kharzeev:2012py}. The reason is that
the calculation in \cite{Kharzeev:2012py} effectively used
$D_{\vr}=e^{-\frac{1}{8}Q_{s}^2 r_\perp^2\ln\frac{1}{\mu r_\perp}}$
instead of eq.~\eqref{eq:mvapp} to calculate dipole gluon
distributions. The expression used in \cite{Kharzeev:2012py} is the
correct expression in the framework of the McLerran-Venugopalan
model. We used the Gaussian form of eq.~\eqref{eq:mvapp} for
convenience to efficiently check how our results reduce to those of
\cite{Kharzeev:2012py}.} to the result of eq.~(27) of Kharzeev et.
al. in \cite{Kharzeev:2012py} once we define the function
$\phi_T(r,z)$ in that paper to be $\phi_T(r, z) = \frac{
|R(0)|\sqrt{\pi}}{\sqrt{m^3} \sqrt{2N_c}}$. When we integrate our
results over $p_\perp$, we recover the result in
ref.~\cite{Dominguez:2011cy} for the total $\jpsi$ cross-section.

We conclude therefore that results for $\jpsi$ differential cross
section in high energy proton-nucleus collisions derived by
Dominguez et.~al. in Refs.~\cite{Dominguez:2011cy,Kharzeev:2012py}
correspond to our color singlet results when we work in the
quasi-classical approximation of the McLerran-Venugopalan model for
the dipole/quadrupole correlators\footnote{Note that the model for
$\jpsi$ wave function in \cite{Dominguez:2011cy,Kharzeev:2012py} is
different from ours. However, using the power counting in NRQCD, one
finds that the difference is suppressed by $v^2$. Thus the
equivalence holds to leading order in $v$ accuracy.}.

We note however, that our expressions [for instance
eq.~\eqref{eq:dscollCS}] allow for  a full JIMWLK treatment of
quarkonium production, including small $x$ evolution and coherent
multiple scattering in a consistent way.  Another advantage of our
formalism is  that we also have color octet contributions which as
we have discussed are important when $p_\perp\geq Q_{s}$.

\subsection{Comparison to the Color Evaporation model}\label{sec:CEM}

The Color Evaporation model (CEM) is often employed in the
literature to study heavy quarkonium production in high energy
proton-nucleus collisions.  For recent work relating the CGC
framework to the CEM, see~\cite{Fujii:2006ab,Fujii:2013gxa}. In this
model, heavy quarkonium production is factorized into two steps: the
perturbative (weak coupling) production of a heavy quark pair with
invariant mass $M$ followed by a non-perturbative hadronization
process. The latter is assumed to have a universal transition
probability for the pair to become a bound quarkonium state. It is
assumed that the transition probability is the same for all heavy
quark pairs whose invariant mass is less than the mass threshold of
producing two open flavor heavy mesons.

Taking $\jpsi$ production as an example, the cross section can be written as
\begin{align}\label{eq:cem}
\frac{d {\sigma}_{\jpsi}}{d^2\vp d y} = F_{\jpsi}
\int_{4m_c^2}^{4m_D^2} d M^2 \frac{d {\sigma}_{c\bar c}}{d M^2d^2\vp
d y},
\end{align}
where $F_{\jpsi}$ is a constant non-perturbative transition probability and is independent of the color and spin of the heavy quark pair, $m_c$ ($M_D$) is the charm quark ($D$-meson) mass, and $M$
is the invariant mass of the charm quark pair.

If we decompose the expression in eq.~\eqref{eq:cem} into color
singlet and color octet contributions, the latter will be larger
than the former by an factor of $N_c^2-1$.  This corresponds to the
ratio of the color states for both contributions. As a result, in
the large $N_c$ limit, only color octet contributions remain in the
CEM. This simple analysis agrees with the explicit calculations in
\cite{Fujii:2006ab,Fujii:2013gxa}. In these papers, the CEM
 expressions for $J/\psi$ production involve only the dipole gluon
distribution. This can be contrasted with our NRQCD framework.  In
our case, while the the color octet channel in eq.~\eqref{eq:dsktCO}
involves only the dipole gluon distribution, the color singlet
channel in eq.~\eqref{eq:dsktCS} involves the quadrupole gluon
distribution as well.

The power counting in NRQCD gives color octet contributions that are
suppressed by $v^4$ relative to the color singlet contributions to
$\jpsi$ production. As $v^4<\frac{1}{N_c^2}$ for both charmonium and
bottomonium states, the color octet contributions are generally less
important than color singlet contribution in this case. Exceptions
exist for special kinematic region ( such as at large $p_\perp$),
where the color octet mechanism may be dominant.  Even so, though
the color octet channels may dominate, the predictions of  NRQCD
factorization and the CEM can be different. This is because NRQCD
factorization assigns a different parameter for each color octet
channel, while the CEM assumes all these parameters to be the same.

\section{Summary and outlook}\label{sec:summary}

The Color Glass Condensate (CGC) is a powerful formalism to systematically compute the final states in deeply inelastic scattering and hadron-hadron scattering experiments at high energies. In proton-nucleus collisions, it allows one to compute both the small $x$ QCD evolution of the projectile and target wavefunctions, as well as  multiple scattering effects due to the large number of color charges in the nuclear target. The CGC formalism was used previously to derive the cross-sections for the production of heavy quark pairs in \cite{Blaizot:2004wv}. However, only the Color Evaporation Model (CEM) was used previously to compute the production of quarkonium bound states~\cite{Fujii:2006ab,Fujii:2013gxa}.

The production of quarkonium bound states can be quantified within the Non-relativistic QCD (NRQCD) framework. The magnitude of long distance color singlet and color octet bound state matrix elements in different spin and angular momentum configurations can be categorized in powers of the relative velocity between the heavy quark-antiquark pair. Further, these universal matrix elements can be determined independently by experimental measurements. The short distance hard partonic cross sections however have to be computed in perturbative QCD.

In this work, we combined for the first time the CGC and NRQCD
formalisms for quarkonium production. The former is used to compute
the short distance matrix elements in weak coupling and the latter
to describe the hadronization of the produced intermediate color
singlet and color octet heavy quark pairs. Interestingly, we find
that the intermediate color states are sensitive to different
universal multi-gluon correlators in high energy QCD. The color
singlet channel is sensitive to the QCD evolution of dipole and
quadrupole Wilson line correlators while the color octet channel is
sensitive to those of the dipole correlators alone. The fact that we
were able to reproduce non-trivial results for color singlet
$J/\psi$ production in a quasi-classical approximation gives us
confidence in the power and validity of our results.

Because the dipole and quadrupole correlators are universal, they
can be measured in other final states (such as inclusive
photon-hadron and di-hadron correlations) in proton-nucleus
collisions, and used to predict the production cross-sections of a
number of quarkonium states. Conversely, the extraction of these
correlators from combinations of production cross-sections of
quarkonium states compared to data, can be used to predict
cross-sections for other final states in high energy proton-nucleus
collisions.

One thus has the possibility to further systematically test and
extend the NRQCD framework, as well as the CGC effective theory
describing the behavior of multi-gluon correlators in hadron
wavefunctions. Understanding these ``cold" nuclear matter
cross-sections then provide a benchmark for the interpretation of
the same in nucleus-nucleus collisions. The recently demonstrated
ability of LHC and RHIC experiments to compare final states in
vastly different systems with the same bulk properties (such as
events with the same number of charged hadrons) make such studies
especially compelling in order to study the transition from cold
matter to hot matter effects in the production of different
quarkonium states.

We have not attempted in this work to perform the numerical
computations necessary to compare our results to those from collider
experiments. This work is numerically challenging (particularly for
the color singlet channel) but feasible. Work in this direction is
in progress and will be reported in the near future.

\begin{acknowledgments}
We would like to thank F. Dominguez, A. Dumitru, K. Dusling, H.
Fujii, J. Lansberg, E. Levin, L. McLerran, Y. Nara, J. Qiu, B. Schenke and F.
Yuan for useful discussions. This work was supported by the U.S.
Department of Energy, under Contract No.~DE-AC52-06NA25396 (ZK) and
DE-AC02-98CH10886 (YM and RV). The Feynman diagrams were drawn using
Jaxodraw~\cite{Binosi:2008ig}.
\end{acknowledgments}


\appendix
\renewcommand{\theequation}{\thesection\arabic{equation}}
\allowdisplaybreaks

\section{NRQCD projectors}\label{sec:projectors}

In this appendix, we list NRQCD projectors for all $S$-wave channels
and $P$-wave channels, which are used to calculate hard part in
appendix~\ref{sec:hard}. For $\cSa$ channels, total angular momentum
equals to its spin angular momentum, we thus need
\begin{align}
\sum_{S_z}\epsilon^{*\alpha}(S_z)
\epsilon^{\alpha'}(S_z)=\mathbb{P}^{\alpha \alpha'},
\end{align}
where $\mathbb{P}^{\alpha \alpha'}\equiv-g^{\alpha
\alpha'}+\frac{p^\alpha p^{\alpha'}}{p^2}$. For $\aPa$ channels,
total angular momentum equals to its orbital angular momentum,  we
thus need
\begin{align}
\sum_{L_z}\epsilon^{*\beta}(L_z)
\epsilon^{\beta'}(L_z)=\mathbb{P}^{\beta \beta'}.
\end{align}
For $\cPj$ channels, using the following notation,
\begin{align}
\epsilon^{*\alpha \beta}(J,J_z)\equiv\sum_{L_z, S_z}\left\langle 1
L_z; 1 S_z|J J_z\right\rangle\epsilon^{*\beta}(L_z)
\epsilon^{*\alpha}(S_z),
\end{align}
we find
\begin{subequations}
\begin{align}
\sum_{J_z}\epsilon^{*\alpha \beta}(0,J_z)\epsilon^{\alpha'
\beta'}(0,J_z)=&\frac{1}{3}\mathbb{P}^{\alpha
\beta}\mathbb{P}^{\alpha'
\beta'},\\
\sum_{J_z}\epsilon^{*\alpha \beta}(1,J_z)\epsilon^{\alpha'
\beta'}(1,J_z)=&\frac{1}{2}\left(\mathbb{P}^{\alpha
\alpha'}\mathbb{P}^{\beta \beta'}-\mathbb{P}^{\alpha
\beta'}\mathbb{P}^{\alpha'
\beta}\right),\\
\sum_{J_z}\epsilon^{*\alpha \beta}(2,J_z)\epsilon^{\alpha'
\beta'}(2,J_z)=&\frac{1}{2}\left(\mathbb{P}^{\alpha
\alpha'}\mathbb{P}^{\beta \beta'}+\mathbb{P}^{\alpha
\beta'}\mathbb{P}^{\alpha'
\beta}\right)-\frac{1}{3}\mathbb{P}^{\alpha
\beta}\mathbb{P}^{\alpha' \beta'}.
\end{align}
\end{subequations}
For $\COcPj$ channels, because of CO LDMEs are related, we sometimes
only need the expression by summing over $J$, which gives
\begin{align}\label{eq:3pjproj}
\sum_{J, J_z}\epsilon^{*\alpha \beta}(J,J_z)\epsilon^{\alpha'
\beta'}(J,J_z)=&\mathbb{P}^{\alpha \alpha'}\mathbb{P}^{\beta
\beta'}.
\end{align}

\section{Calculation of the hard part}\label{sec:hard}

In this appendix, we give results of hard part for all $S$-wave
channels and $P$-wave channels. These results are sufficient for
phenomenological study of common heavy quarkonia production in pA
collision using NRQCD factorization.

\subsection{Hard part for color singlet channels}\label{sec:cshard}

\subsubsection{Complete results}

To calculate $\Gamma_1^\kappa$ defined in eq.~\eqref{eq:defgamma1},
we first calculate the following quantities
\begin{align}
W^\kappa\equiv\overline{\sum_{J_z}} \mathcal{F}^{\kappa,
J_z}_{q\bar{q}}\left(p,\vka,\vk\right) \mathcal{F}^{\dagger\kappa,
J_z}_{q\bar{q}}\left(p,\vka,\vkp\right).
\end{align}
We find
\begin{subequations}\label{eq:gamma1}
\begin{align}
\begin{split}\label{eq:gamma3s11}
W^{\CScSa}=&\frac{k_{1\perp}^2\left( k_{1\perp}^2 + 4m^2\right)}{6 m
X_{l_\perp} X_{l'_\perp}}+\cdots,
\end{split}\\
\begin{split}
W^{\CScPz}=&\frac{2\vka\cdot\vl \vka\cdot\vlp}{3m^3 X_{l_\perp}
X_{l'_\perp}}+\frac{\vka\cdot\vlp \left[k_{1\perp}^2
(\vp-\vka)\cdot\vl +4m^2\vka\cdot\vl\right]}{3m^3 X_{l_\perp}^2
X_{l'_\perp}}\\
&+\frac{\vka\cdot\vl \left[k_{1\perp}^2 (\vp-\vka)\cdot\vlp
+4m^2\vka\cdot\vlp\right]}{3m^3 X_{l_\perp}
X_{l'_\perp}^2}\\
&+\frac{\left[k_{1\perp}^2 (\vp-\vka)\cdot\vl
+4m^2\vka\cdot\vl\right] \left[k_{1\perp}^2 (\vp-\vka)\cdot\vlp
+4m^2\vka\cdot\vlp\right]}{6m^3 X_{l_\perp}^2 X_{l'_\perp}^2},
\end{split}\\
\begin{split}
W^{\CScPa}=&\frac{4\left(k_{1\perp}^2 \vl\cdot\vlp - \vka\cdot\vl
\vka\cdot\vlp\right)}{3m^3 }\left(\frac{1}{X_{l_\perp}
X_{l'_\perp}}-\frac{m^2}{X_{l_\perp}^2
X_{l'_\perp}}-\frac{m^2}{X_{l_\perp} X_{l'_\perp}^2}\right) \\
&+ \frac{1}{3m X_{l_\perp}^2 X_{l'_\perp}^2}\left[k_{1\perp}^4
\vl\cdot\vlp+k_{1\perp}^2 \left(3\vka\cdot\vl\vka\cdot\vlp
-2\vka\cdot\vl\vp\cdot\vlp\right.\right.\\
&\left.\left. -2\vka\cdot\vlp\vp\cdot\vl +4m^2 \vl\cdot\vlp+
\vp\cdot\vl \vp\cdot\vlp \right)
-4m^2\vka\cdot\vl\vka\cdot\vlp\right],
\end{split}\\
\begin{split}
W^{\CScPb}=&\frac{4\vka\cdot\vl \vka\cdot\vlp}{15m^3X_{l_\perp}
X_{l'_\perp} } + \frac{2\vka\cdot\vlp \left[k_{1\perp}^2
(\vp-\vka)\cdot\vl -2m^2\vka\cdot\vl\right]}{15m^3X_{l_\perp}^2
X_{l'_\perp} }\\
& +  \frac{2\vka\cdot\vl \left[k_{1\perp}^2 (\vp-\vka)\cdot\vlp
-2m^2\vka\cdot\vlp\right]}{15m^3X_{l_\perp} X_{l'_\perp}^2 } +
\frac{1}{15m^3X_{l_\perp}^2 X_{l'_\perp}^2}\\
& \times \left\{ k_{1\perp}^4
\left[(\vp-\vka)\cdot\vl(\vp-\vka)\cdot\vlp+3m^2
\vl\cdot\vlp\right]\right.\\
&+k_{1\perp}^2 m^2
\left(\vka\cdot\vl\vka\cdot\vlp-2\vka\cdot\vl\vp\cdot\vlp
-2\vka\cdot\vlp\vp\cdot\vl \right.\\
&\left. \left. +12m^2 \vl\cdot\vlp+ 3\vp\cdot\vl \vp\cdot\vlp
\right) + 4m^4\vka\cdot\vl \vka\cdot\vlp \right\},
\end{split}\\
\begin{split}
W^{\CSaSz}=&\frac{2\left(k_{1\perp}^2 \vl\cdot\vlp - \vka\cdot\vl
\vka\cdot\vlp\right)}{m X_{l_\perp} X_{l'_\perp}},
\end{split}\\
\begin{split}
W^{\CSaPa}=&\frac{\left(\vka\cdot\vp\right)^2-k_{1\perp}^2
p_\perp^2}{6m^3 X_{l_\perp} X_{l'_\perp}}+\frac{\left(
\vp-\vka\right)\cdot\vl \left(k_{1\perp}^2 \vp\cdot\vl -
\vka\cdot\vl \vka\cdot\vp\right)}{3m^3 X_{l_\perp}^2
X_{l'_\perp}}\\
&+\frac{\left( \vp-\vka\right)\cdot\vlp \left(k_{1\perp}^2
\vp\cdot\vlp - \vka\cdot\vlp \vka\cdot\vp\right)}{3m^3 X_{l_\perp}
X_{l'_\perp}^2}-\frac{2}{3m^3 X_{l_\perp}^2 X_{l'_\perp}^2}\\
&\times \left(k_{1\perp}^2 \vl\cdot\vlp - \vka\cdot\vl
\vka\cdot\vlp\right)
\left[(\vp-\vka)\cdot\vl(\vp-\vka)\cdot\vlp+4m^2
\vl\cdot\vlp\right],
\end{split}
\end{align}
\end{subequations}
where
\begin{align}
\begin{split}
\vl=\vk-\frac{\vp-\vka}{2},\quad \vlp=\vkp-\frac{\vp-\vka}{2},
\end{split}
\end{align}
and
\begin{align}
\begin{split}
X_{l_\perp}=l_{\perp}^2+\frac{k_{1\perp}^2}{4}+m^2,\quad
X_{l'_\perp}={l'}_{\perp}^2+\frac{k_{1\perp}^2}{4}+m^2.
\end{split}
\end{align}
The ``$\cdots$'' in $W^{\CScSa}$ represents terms that are
independent of either $\vl$ or $\vlp$, which will eventually
contribute to $\Gamma_1^{\CScSa}$ in terms of $\delta(\vr)$ or
$\delta(\vrp)$. Let us denote the following abbreviations
\begin{subequations}
\begin{align}
Z_0\equiv&\frac{1}{2\pi} \underset{\vk}{\int} \frac{ e^{i
\vl\cdot\vr}}{X_{l_\perp}}=K_0\left(r_\perp
\sqrt{\frac{k_{1\perp}^2}{4}+m^2}\right),\\
Z_1\equiv&\frac{1}{2\pi} \underset{\vk}{\int} \frac{ e^{i
\vl\cdot\vr}}{X_{l_\perp}^2}=\frac{r_\perp}{
2\sqrt{\frac{k_{1\perp}^2}{4}+m^2}}K_1\left(r_\perp
\sqrt{\frac{k_{1\perp}^2}{4}+m^2}\right),\\
Z'_0\equiv&\frac{1}{2\pi} \underset{\vkp}{\int} \frac{ e^{-i
\vlp\cdot\vrp}}{X_{l'_\perp}}=K_0\left(r'_\perp
\sqrt{\frac{k_{1\perp}^2}{4}+m^2}\right),\\
Z'_1\equiv&\frac{1}{2\pi} \underset{\vkp}{\int} \frac{ e^{-i
\vlp\cdot\vrp}}{X_{l'_\perp}^2}=\frac{r'_\perp}{
2\sqrt{\frac{k_{1\perp}^2}{4}+m^2}}K_1\left(r'_\perp
\sqrt{\frac{k_{1\perp}^2}{4}+m^2}\right),
\end{align}
\end{subequations}
where $K_{0,1}$ are the modified Bessel functions.
Then, $\Gamma_1^\kappa$ can be obtained by
\begin{align}\label{eq:GammaWCS}
\Gamma_1^\kappa = \frac{1}{(2\pi)^2} \underset{\vk, \vkp}{\int}
e^{i \vl\cdot\vr}e^{-i
\vlp\cdot\vrp} W^\kappa.
\end{align}
For $\kappa=\CScSa$, we obtain $\Gamma^{\kappa}_1$ from $W^{\kappa}$ by
doing the replacement
\begin{align}
X_{l_\perp}^{-1} \to Z_0, \quad X_{l'_\perp}^{-1} \to Z'_0.
\end{align}
For $\kappa=\CScPz, \CScPa, \CScPb, \CSaSz$, we obtain
$\Gamma^{\kappa}_1$ from $W^{\kappa}$ by doing the replacement
\begin{align}
\begin{split}
&X_{l_\perp}^{-2} \to 2 \frac{\partial Z_1}{\partial r_\perp^2} ,
\quad X_{l_\perp}^{-1} \to 2\frac{\partial
Z_0}{\partial r_\perp^2} ,\\
&X_{l'_\perp}^{-2} \to 2\frac{\partial Z'_1}{\partial {r'}_\perp^2}
, \quad X_{l'_\perp}^{-1} \to 2\frac{\partial
Z'_0}{\partial {r'}_\perp^2} ,\\
&\vl\to\vr,\quad \vlp\to\vrp.
\end{split}
\end{align}
For $\kappa=\CSaPa$, we obtain
\begin{align}
\begin{split}
\Gamma_1^{\CSaPa}=&-\left(Z_0+4 \frac{\partial Z_1}{\partial
{r}_\perp^2}\right)\left(Z'_0+4\frac{\partial Z'_1}{\partial
{r'}_\perp^2}\right)\frac{k_{1\perp}^2
p_\perp^2-\left(\vka\cdot\vp\right)^2}{6m^3}\\
&-4 \frac{\partial^2 Z_1}{\partial^2
r_\perp^2}\left(Z'_0+4\frac{\partial Z'_1}{\partial
{r'}_\perp^2}\right) \frac{\left( \vp-\vka\right)\cdot\vr
\left(k_{1\perp}^2 \vp\cdot\vr - \vka\cdot\vr
\vka\cdot\vp\right)}{3m^3}\\
&-4 \frac{\partial^2 Z'_1}{\partial^2 {r'}_\perp^2}\left(Z_0+4
\frac{\partial Z_1}{\partial {r}_\perp^2}\right) \frac{\left(
\vp-\vka\right)\cdot\vrp \left(k_{1\perp}^2 \vp\cdot\vrp -
\vka\cdot\vrp \vka\cdot\vp\right)}{3m^3}\\
&-16 \frac{\partial Z_1}{\partial r_\perp^2} \frac{\partial
Z'_1}{\partial {r'}_\perp^2} \frac{2 k_{1\perp}^2}{3m}-16
\frac{\partial^2 Z_1}{\partial^2 r_\perp^2} \frac{\partial
Z'_1}{\partial {r'}_\perp^2} \frac{4}{3m}\left[k_{1\perp}^2
r_\perp^2-\left(\vka\cdot\vr\right)^2\right]\\
&-16 \frac{\partial^2 Z'_1}{\partial^2 {r'}_\perp^2} \frac{\partial
Z_1}{\partial {r}_\perp^2} \frac{4}{3m}\left[k_{1\perp}^2
{r'}_\perp^2-\left(\vka\cdot\vrp\right)^2\right]-16 \frac{\partial^2
Z_1}{\partial^2 r_\perp^2} \frac{\partial^2 Z'_1}{\partial^2
{r'}_\perp^2}
\frac{2}{3m^3}\\
&\times \left(k_{1\perp}^2 \vr\cdot\vrp - \vka\cdot\vr
\vka\cdot\vrp\right)
\left[(\vp-\vka)\cdot\vr(\vp-\vka)\cdot\vrp+4m^2
\vr\cdot\vrp\right].
\end{split}
\end{align}

\subsubsection{Collinear limit}

Define
\begin{align}
\tilde{W}^\kappa \equiv \underset{k_{1\perp}\to 0}{\text{lim.}}
\frac{1}{2\pi}\int_0^{2\pi} d\theta_1 \frac{W^\kappa}{k_{1\perp}^2},
\end{align}
we find
\begin{subequations}\label{eq:gamma1coll}
\begin{align}
\begin{split}
\tilde{W}^{\CScSa}=&\frac{2m}{3 \tilde{X}_{l_\perp}
\tilde{X}_{l'_\perp}} + \cdots,
\end{split}\\
\begin{split}
\tilde{W}^{\CScPz}=&\frac{\vl\cdot\vlp}{3m^3}
\left(\frac{1}{\tilde{X}_{l_\perp} \tilde{X}_{l'_\perp}}+
\frac{2m^2}{\tilde{X}_{l_\perp}^2 \tilde{X}_{l'_\perp}} +
\frac{2m^2}{\tilde{X}_{l_\perp} \tilde{X}_{l'_\perp}^2}+
\frac{4m^4}{\tilde{X}_{l_\perp}^2 \tilde{X}_{l'_\perp}^2}\right),
\end{split}\\
\begin{split}
\tilde{W}^{\CScPa}=&\frac{2\vl\cdot\vlp}{3m^3}
\left(\frac{1}{\tilde{X}_{l_\perp} \tilde{X}_{l'_\perp}}
-\frac{m^2}{\tilde{X}_{l_\perp}^2 \tilde{X}_{l'_\perp}}
-\frac{m^2}{\tilde{X}_{l_\perp} \tilde{X}_{l'_\perp}^2}+
\frac{m^4+\frac{m^2}{2}\frac{\vp\cdot\vl \vp\cdot\vlp
}{\vl\cdot\vlp}}{\tilde{X}_{l_\perp}^2
\tilde{X}_{l'_\perp}^2}\right),
\end{split}\\
\begin{split}
\tilde{W}^{\CScPb}=&\frac{2\vl\cdot\vlp}{15m^3}
\left(\frac{1}{\tilde{X}_{l_\perp} \tilde{X}_{l'_\perp}}
-\frac{m^2}{\tilde{X}_{l_\perp}^2 \tilde{X}_{l'_\perp}}
-\frac{m^2}{\tilde{X}_{l_\perp} \tilde{X}_{l'_\perp}^2}+
\frac{7m^4+\frac{3m^2}{2}\frac{\vp\cdot\vl \vp\cdot\vlp
}{\vl\cdot\vlp}}{\tilde{X}_{l_\perp}^2
\tilde{X}_{l'_\perp}^2}\right),
\end{split}\\
\begin{split}
\tilde{W}^{\CSaSz}=& \frac{\vl\cdot\vlp}{m \tilde{X}_{l_\perp}
\tilde{X}_{l'_\perp}},
\end{split}\\
\begin{split}
\tilde{W}^{\CSaPa}=&\frac{1}{12m^3}
\left[-\frac{p_\perp^2}{\tilde{X}_{l_\perp} \tilde{X}_{l'_\perp}}
+\frac{2\left(\vp\cdot\vl\right)^2}{\tilde{X}_{l_\perp}^2
\tilde{X}_{l'_\perp}}
+\frac{2\left(\vp\cdot\vlp\right)^2}{\tilde{X}_{l_\perp}
\tilde{X}_{l'_\perp}^2}- \frac{4(\vl\cdot \vlp)^2
\left(4m^2+\frac{\vp\cdot\vl \vp\cdot\vlp
}{\vl\cdot\vlp}\right)}{\tilde{X}_{l_\perp}^2
\tilde{X}_{l'_\perp}^2}\right],
\end{split}
\end{align}
\end{subequations}
where
\begin{align}
\begin{split}
\tilde{X}_{l_\perp}=l_{\perp}^2+m^2,\quad
\tilde{X}_{l'_\perp}={l'}_{\perp}^2+m^2.
\end{split}
\end{align}
Similarly, the ``$\cdots$'' in $\tilde{W}^{\CScSa}$ represents terms
that are independent of either $\vl$ or $\vlp$, which has no
contribution for cross section. Let us denote the following
abbreviations
\begin{align}
\tilde{Z}_0\equiv&K_0\left(r_\perp m\right),\quad
\tilde{Z}_1\equiv&\frac{r_\perp}{ 2m}K_1\left(r_\perp m\right),\quad
\tilde{Z}'_0\equiv&K_0\left(r'_\perp m\right),\quad
\tilde{Z}'_1\equiv&\frac{r_\perp'}{ 2m}K_1\left(r'_\perp m\right).
\end{align}
Then $\tilde{\Gamma}_1^\kappa$ defined in eq.~\eqref{cs-limit} can be obtained by
\begin{align}
\tilde{\Gamma}_1^\kappa = \frac{1}{(2\pi)^2} \underset{\vk, \vkp}{\int}
e^{i \vl\cdot\vr}e^{-i
\vlp\cdot\vrp} \tilde{W}^\kappa.
\end{align}
For $\kappa=\CScSa$, we obtain $\tilde{\Gamma}^{\kappa}_1$ from
$\tilde{W}^{\kappa}$ by doing the replacement
\begin{align}
\tilde{X}_{l_\perp}^{-1} \to \tilde{Z}_0, \quad
\tilde{X}_{l'_\perp}^{-1} \to \tilde{Z}'_0.
\end{align}
For $\kappa=\CScPz, \CScPa, \CScPb, \CSaSz$, we obtain
$\tilde{\Gamma}^{\kappa}_1$ from $\tilde{W}^{\kappa}$ by doing the
replacement
\begin{align}
\begin{split}
&\tilde{X}_{l_\perp}^{-2} \to 2 \frac{\partial \tilde{Z}_1}{\partial
r_\perp^2} , \quad \tilde{X}_{l_\perp}^{-1} \to 2\frac{\partial
\tilde{Z}_0}{\partial r_\perp^2} ,\\
&\tilde{X}_{l'_\perp}^{-2} \to 2\frac{\partial
\tilde{Z}'_1}{\partial {r'}_\perp^2} , \quad
\tilde{X}_{l'_\perp}^{-1} \to 2\frac{\partial
\tilde{Z}'_0}{\partial {r'}_\perp^2} ,\\
&\vl\to\vr,\quad \vlp\to\vrp.
\end{split}
\end{align}
For $\kappa=\CSaPa$, we obtain
\begin{align}
\begin{split}
\tilde{\Gamma}_1^{\CSaPa}=&-\frac{p_\perp^2}{12m^3}\left[\left(\tilde{Z}_0+4
\frac{\partial \tilde{Z}_1}{\partial
{r}_\perp^2}\right)\left(\tilde{Z}'_0+4\frac{\partial
\tilde{Z}'_1}{\partial
{r'}_\perp^2}\right)\right.\\
&+4 \frac{\partial^2 \tilde{Z}_1}{\partial^2
r_\perp^2}\left(\tilde{Z}'_0+4\frac{\partial \tilde{Z}'_1}{\partial
{r'}_\perp^2}\right) \frac{2(\vp\cdot\vr)^2}{p_\perp^2} +4
\frac{\partial^2 \tilde{Z}'_1}{\partial^2
{r'}_\perp^2}\left(\tilde{Z}_0+4
\frac{\partial \tilde{Z}_1}{\partial {r}_\perp^2}\right) \frac{2(\vp\cdot\vrp)^2}{p_\perp^2}\\
&+16 \left( \frac{\partial \tilde{Z}_1}{\partial r_\perp^2}
\frac{\partial \tilde{Z}'_1}{\partial {r'}_\perp^2} +
{r}_\perp^2\frac{\partial^2 \tilde{Z}_1}{\partial^2 r_\perp^2}
\frac{\partial \tilde{Z}'_1}{\partial {r'}_\perp^2} +{r'}_\perp^2
\frac{\partial^2 \tilde{Z}'_1}{\partial^2 {r'}_\perp^2}
\frac{\partial \tilde{Z}_1}{\partial {r}_\perp^2} \right)
\frac{8m^2 }{p_\perp^2}\\
&\left.+16 \frac{\partial^2 \tilde{Z}_1}{\partial^2 r_\perp^2}
\frac{\partial^2 \tilde{Z}'_1}{\partial^2 {r'}_\perp^2}
\frac{4\left(\vr\cdot\vrp\right)^2}{p_\perp^2}\left(4m^2+\frac{\vp\cdot\vr\vp\cdot\vrp}{\vr\cdot\vrp}\right)\right].
\end{split}
\end{align}

\subsection{Hard part for color octet channels}\label{sec:cohard}

\subsubsection{Complete results}

From the definition for $\Gamma_8^{\kappa}$ in
eq.~\eqref{eq:defgamma8}, we get
\begin{subequations}\label{eq:gamma8}
\begin{align}
\begin{split}
\Gamma_8^{\COcSa}=&\frac{2k_{1\perp}^2\left[\left(\vp-\vka\right)^2
+ 4m^2\right]}{3 m^3
(p_\perp^2+4m^2)}-\frac{4k_{1\perp}^2\left[\left(\vp-\vka\right)^2+\vka\cdot\vp
+ 4m^2\right]}{3 m X_{l_\perp}
(p_\perp^2+4m^2)}\\
&+\frac{k_{1\perp}^2\left( k_{1\perp}^2 + 4m^2\right)}{6 m
X_{l_\perp}^2},
\end{split}\\
\begin{split}
\Gamma_8^{\COcPz}=&\frac{2(\vka\cdot\vl)^2}{3m^3
X^2_{l_\perp}}+\frac{2\vka\cdot\vl \left[k_{1\perp}^2
(\vp-\vka)\cdot\vl +4m^2\vka\cdot\vl\right]}{3m^3 X_{l_\perp}^3}\\
&+\frac{\left[k_{1\perp}^2 (\vp-\vka)\cdot\vl
+4m^2\vka\cdot\vl\right]^2}{6m^3 X_{l_\perp}^4},
\end{split}\\
\begin{split}
\Gamma_8^{\COcPa}=&\frac{4\left[k_{1\perp}^2 l_{\perp}^2 -
(\vka\cdot\vl)^2\right]}{3m^3 }\left(\frac{1}{X_{l_\perp}^2 }
-\frac{2m^2}{X_{l_\perp}^3}\right)+ \frac{1}{3m
X_{l_\perp}^4}\left\{k_{1\perp}^4
l_{\perp}^2 \right.\\
&\left.+k_{1\perp}^2
\left[(\vp-\vka)\cdot\vl(\vp-3\vka)\cdot\vl+4m^2 l_{\perp}^2 \right]
-4m^2(\vka\cdot\vl)^2\right\},
\end{split}\\
\begin{split}
\Gamma_8^{\COcPb}=&\frac{4(\vka\cdot\vl)^2}{15m^3X_{l_\perp}^2} +
\frac{4\vka\cdot\vl \left[k_{1\perp}^2 (\vp-\vka)\cdot\vl
-2m^2\vka\cdot\vl\right]}{15m^3X_{l_\perp}^3}\\
& + \frac{1}{15m^3X_{l_\perp}^4}\left\{ k_{1\perp}^4
\left[\left((\vp-\vka)\cdot\vl\right)^2+3m^2
l_{\perp}^2\right]\right.\\
&\left.+k_{1\perp}^2 m^2
\left[(\vp-\vka)\cdot\vl(3\vp-\vka)\cdot\vl+12m^2 l_{\perp}^2
\right]+ 4m^4(\vka\cdot\vl)^2 \right\},
\end{split}\\
\begin{split}
\Gamma_8^{\COcPj}=&\frac{4k_{1\perp}^2 l_{\perp}^2 -
2(\vka\cdot\vl)^2}{9m^3 X_{l_\perp}^2}+ \frac{2k_{1\perp}^2
\vka\cdot\vl(\vp-\vka)\cdot\vl
-8m^2\left[k_{1\perp}^2l_{\perp}^2-(\vka\cdot\vl)^2\right]}{9m^3 X_{l_\perp}^3} \\
&+ \frac{k_{1\perp}^2 (k_{1\perp}^2+4m^2)\left\{\left[
(\vp-\vka)\cdot\vl\right]^2 +4m^2l_{\perp}^2\right\} }{18m^3
X_{l_\perp}^4},
\end{split}\\
\begin{split}
\Gamma_8^{\COaSz}=&\frac{2\left[k_{1\perp}^2 l_{\perp}^2 -
(\vka\cdot\vl)^2\right]}{m X_{l_\perp}^2},
\end{split}\\
\begin{split}
\Gamma_8^{\COaPa}=&\frac{\left(\vka\cdot\vp\right)^2-k_{1\perp}^2
p_\perp^2}{6m^3 X_{l_\perp}^2}+\frac{2\left( \vp-\vka\right)\cdot\vl
\left(k_{1\perp}^2 \vp\cdot\vl - \vka\cdot\vl
\vka\cdot\vp\right)}{3m^3 X_{l_\perp}^3}\\
&-\frac{2 \left[k_{1\perp}^2 l_{\perp}^2 - (\vka\cdot\vl)^2\right]
\left\{\left[(\vp-\vka)\cdot\vl\right]^2+4m^2
l_{\perp}^2\right\}}{3m^3 X_{l_\perp}^4} ,
\end{split}
\end{align}
\end{subequations}
where $\Gamma_8^{\COcPj}$ is obtained using the projector in
eq.~\eqref{eq:3pjproj}. It is easy to find that
$\Gamma_8^{\COcPz}+3\Gamma_8^{\COcPa}+5\Gamma_8^{\COcPb}=9\Gamma_8^{\COcPj}$.

\subsubsection{Collinear limit}

In the collinear limit of proton side, we get the results for
$\tilde{\Gamma}_8^{\kappa}$ defined in eq.~\eqref{co-limit},
\begin{subequations}\label{eq:gamma8coll}
\begin{align}
\begin{split}
\tilde{\Gamma}_8^{\COcSa}=&\frac{2}{3
m^3}\left(1-\frac{m^2}{\tilde{X}_{l_\perp}}\right)^2,
\end{split}\\
\begin{split}
\tilde{\Gamma}_8^{\COcPz}=&\frac{l_\perp^2}{3m^3
\tilde{X}^2_{l_\perp}}\left(1+\frac{2m^2}{\tilde{X}_{l_\perp}}\right)^2,
\end{split}\\
\begin{split}
\tilde{\Gamma}_8^{\COcPa}=&\frac{2l_\perp^2}{3m^3
\tilde{X}^2_{l_\perp}}\left[\left(1-\frac{m^2}{\tilde{X}_{l_\perp}}\right)^2+
\frac{m^2(\vp\cdot\vl)^2}{2l_\perp^2\tilde{X}_{l_\perp}^2}\right],
\end{split}\\
\begin{split}
\tilde{\Gamma}_8^{\COcPb}=&\frac{2l_\perp^2}{15m^3
\tilde{X}^2_{l_\perp}}\left[1-\frac{2m^2}{\tilde{X}_{l_\perp}}+\frac{m^2}{2}
\frac{14m^2+3\frac{(\vp\cdot\vl)^2}{l_\perp^2}}{\tilde{X}_{l_\perp}^2}\right],
\end{split}\\
\begin{split}
\tilde{\Gamma}_8^{\COcPj}=&\frac{l_\perp^2}{3m^3
\tilde{X}^2_{l_\perp}}\left[1-\frac{4m^2}{3\tilde{X}_{l_\perp}}+
\frac{2m^2}{3}
\frac{4m^2+\frac{(\vp\cdot\vl)^2}{l_\perp^2}}{\tilde{X}_{l_\perp}^2}\right],
\end{split}\\
\begin{split}
\tilde{\Gamma}_8^{\COaSz}=&\frac{l_{\perp}^2 }{m
\tilde{X}_{l_\perp}^2},
\end{split}\\
\begin{split}
\tilde{\Gamma}_8^{\COaPa}=&-\frac{1}{12m^3
\tilde{X}^2_{l_\perp}}\left\{p_\perp^2-\frac{4(\vp\cdot\vl)^2}{\tilde{X}_{l_\perp}}+
\frac{4l_\perp^4\left[4m^2+\frac{(\vp\cdot\vl)^2}{l_\perp^2}\right]}{\tilde{X}_{l_\perp}^2}\right\}.
\end{split}
\end{align}
\end{subequations}
Again, we have
$\tilde{\Gamma}_8^{\COcPz}+3\tilde{\Gamma}_8^{\COcPa}+
5\tilde{\Gamma}_8^{\COcPb}=9\tilde{\Gamma}_8^{\COcPj}$.

\bibliographystyle{utphys}
\bibliography{ref}

\end{document}